\documentclass[]{article}

\usepackage{gensymb}
\usepackage[numbers]{natbib}
\usepackage{subcaption}
\usepackage{amsmath}
\usepackage{amssymb}
\usepackage{braket}
\usepackage{float}
\usepackage{longtable}
\usepackage[document]{ragged2e}
\usepackage{graphicx,color}
\graphicspath{{images/}}
\usepackage{rotating}
\usepackage{epstopdf}
\usepackage{comment}
\usepackage[toc,page]{appendix}
\usepackage{epigraph}
\usepackage{multicol}
\usepackage{tabu}
\usepackage[british]{babel}
\usepackage{enumitem}
\usepackage{setspace}
\usepackage{authblk}
\usepackage{authoraftertitle}
\usepackage[a4paper, total={6.2in, 8in}]{geometry}

\begin{document}

\title{\textbf{Novel permanent magnet array geometries for scalable trapped-ion quantum computing in a laser-free entanglement architecture}}

\author{Mitchell G. Peaks$^*$}
\affil{Department of Physics and Astronomy, University of Sussex, Brighton BN1 9QH, United Kingdom}
\affil{$^*$Current affiliation: Duke Quantum Center, Department of Electrical and Computer Engineering and Department of Physics, Duke University, Durham, NC 27708, USA}
\date{email: mitchell.peaks@duke.edu} 

\maketitle

\begin{abstract}
\justifying
A novel design is presented for a permanent magnet array to address specific challenges with scalable trapped-ion quantum computing systems.  Design and optimization of this magnet geometry is motivated by concepts for large-scale Quantum Charge-Coupled Device (QCCD) architectures.  This proposal is relevant to magnetic field gradient schemes for laser-free entanglement using long-wavelength radiation, and individual addressing based on spatially dependent, magnetic field sensitive qubits.  This configuration generates a localized, asymmetric magnetic field, yielding a region for ion transport into and out of a strong magnetic field gradient, while minimizing the absolute field experienced by the ion.  This is a distinct improvement for scalability over dipolar magnet geometries where a strong magnetic field surrounds a magnetic field nil in three dimensions, which is problematic for ion transport applications.  The design also relaxes the alignment constraints for experimental setup by allowing greater tolerance to misalignment in two dimensions.  Additionally, the potential to scale a permanent magnet scheme in QCCD systems circumvents engineering challenges associated with using large electrical currents to generate the field gradient.  Finally, a conceptual discussion is given for incorporating the design into a scalable QCCD type architecture.
\end{abstract}

\section{Introduction}
\justifying

Scalability is essential for any quantum computing paradigm to succeed in eventually running more complex algorithms and useful calculations, often requiring billions of operations \cite{Gidney2021, Kivlichan2020, Reiher2017}.  In addition, Quantum Error Correction (QEC) schemes rely on appreciable scale devices in terms of the number of physical qubits, and with limited error tolerance in order to be effective \cite{Zhao2022, Krinner2022, Ryan2021, Egan2021}.  There are multiple possible candidate technologies for large-scale quantum computing, including trapped-ions \cite{Bruzewicz2019Review}, neutral atoms \cite{Henriet2020quantumcomputing} and superconducting circuits \cite{Kjaergaard2020}.  A promising scheme for building a large-scale quantum computing architecture with trapped-ion qubits is the Quantum Charge-Coupled Device (QCCD) architecture.  In these systems, ions are trapped in an array or grid of traps, typically 2-D, planar traps above which the ions are confined in a RF pseudo-potential, and shuttled via the application of voltages to DC electrodes fabricated into the surface trap structure \cite{Wineland1998, Monroe2013, Lekitsch2015}.  Many such systems seek to leverage the stable electronic energy states arising due to hyperfine structure in atomic ions with non-zero nuclear spin.  These states can have long coherence times, making them desirable qubits for high-fidelity quantum logic and long-lived quantum memory \cite{Harty2014, Ballance2016, Wang2021}.  The frequency splitting between these states is typically in the microwave spectrum, $\approx 3 - 13$~GHz, dependent on the atomic species.  As a result, state transitions can be induced by the direct application of a microwave frequency magnetic field resonant with the transition frequency.  This method is convenient for single qubit transitions, however, the microwave frequency photons do not have sufficient momentum to induce strong coupling to the vibrational modes of shared oscillations between ions, required for generating entanglement for multi-qubit gates.  For this reason, the use of stimulated Raman transitions is common, coupling two hyperfine states via an auxiliary state using a two-photon process.  The population of the auxiliary state is suppressed by adiabatic elimination through a detuning from the transition resonance \cite{Monroe_Raman1995,Wineland1997,Heinzen1990}.  Schemes of this type have been successful for high-fidelity control in experiments with up to tens of ion strings \cite{Debnath2016}, however scaling to larger systems at potentially arbitrary scales will require significant laser-beam power, control and calibration overheads.  Gates implemented using stimulated-Raman transitions  are also fundamentally limited by spontaneous emission \cite{Ozeri2007}.  Another method of performing two-qubit gates is to apply direct microwave control, but with a scheme to modify the coupling such that the effective Lamb-Dicke parameter becomes appreciable; on the order of optical frequency photon coupling.  This has been demonstrated in multiple experiments using a magnetic field gradient to provide a state-dependent force to increase the spin-motion coupling, allowing high-fidelity two-qubit entangling gate operations \cite{Mintert2001, Ospelkaus2008, Lake2015, Srinivas2021}.  Previous experiments using permanent magnets to generate the state-dependent, spin-motion coupling in this manner have relied on dipole or quadrupole magnet configurations with large magnets placed either beside or under the ion trap \cite{Mintert2001, Lake2015, Kawai2016}.  This is a useful configuration for proof-of-principle experiments, as has been demonstrated in several experiments \cite{Johanning2009, Weidt2016}.  However, permanent magnet configurations of this type have difficulties when scaling to larger qubit numbers, particularly when considering an arbitrary scale, modular design architectures, such as that of Lekitsch et al. \cite{Lekitsch2015}.  In this type of architecture, known as a Quantum Charge-Coupled Device (QCCD), the method of transferring quantum information throughout a larger-scale device is conceived to take place via physical ion shuttling \cite{Hucul2008, Kaushal2020}.  Shuttling ions into and out of the designated gate region requires moving the ion through a strong magnetic field in three dimensions, inducing a Lorentz force on the moving ion, causing additional micro-motion dependent upon the confining potential in the axis of the force.  A basic estimate of the force perpendicular to the direction of ion motion can be calculated by applying the Lorentz force law for a single $^{171}$Yb$^+$ ion moving at a typical ion shuttling speed of $1.6$~ms$^{-1}$, for proof-of-principle experiments in scalable QCCD devices \cite{Akhtar2022}.  If the magnetic field is taken as $250$~mT, from a conservative assumption about the maximum field in the ion shuttling path for previous experiments \cite{Peaks2023}, these parameters yield a force of magnitude $\approx6.4\times10^{-20}$~N, and an acceleration of  $\approx 2.3\times10^5$~ms$^{-2}$, with the force vector rotating as the ion passes over different components of the magnet geometry both systematically, due to design, and unsystematically, due to physical defects and misalignment.  While this is small compared with the force induced by a typical confining potential $\approx 10^{-16}$~N radial and $\approx 10^{-19}$~N axial, it is sufficient to perturb the ion motion in the strong, non-linear magnetic field; a source of qubit decoherence.  In addition, moving the qubit through a strong, non-uniform magnetic field allows magnetic field sensitive states to be exposed to additional magnetic field noise.  All of these effects can cause the qubit to decohere and are detrimental to the extended function of a larger-scale quantum computing device.  Phase rotations on the qubit subspace are also caused by exposure of the ion to a changing magnetic field.  This can be problematic to continued, high-fidelity coherent operations if the phase change cannot be tracked systematically.  Design architectures with fixed qubit locations featuring photonic interconnects may be viable with these magnet configurations, however, these have distinct challenges and are outside the scope of this paper \cite{Monroe2014, Stephenson2020}.

\section{General design principles}
	\label{sec:GeneralDesign}
	The magnetic field gradient generation for the novel design is based on a Halbach array, a permanent magnet configuration used in various applications that can produce an asymmetric amplification and diminution of the magnetic flux density surrounding the magnets.  By taking this basic premise, it was possible to design a configuration of magnets to allow a highly localized, strong magnetic field gradient to be produced in a desired axis, chosen based on a principal ion trapping axis along which quantum gate operations are applied.  In this analysis, the axial motion is chosen, this is the axis of weakest confinement for a linear, planar ion trap.  Designing the array in a linear configuration, symmetric about the axial trap direction allows negligible magnetic flux density in the transverse direction, perpendicular to the axial modes in the plane of the ion trap. A linear Halbach array design features a weak and strong magnetic field side, where a strong magnetic field gradient is apparent at the interface between the negligible field region and the edge of the magnets, where the lines of flux transition from sparse to high density in a compressed axial distance.  All simulations herein were generated using the COMSOL Multiphysics $\circledR$ software package, AC/DC module; a numerical modeling tool employing a Finite Element Method (FEM).
	
\begin{figure}[H]
	\centering
	\includegraphics[width=\linewidth]{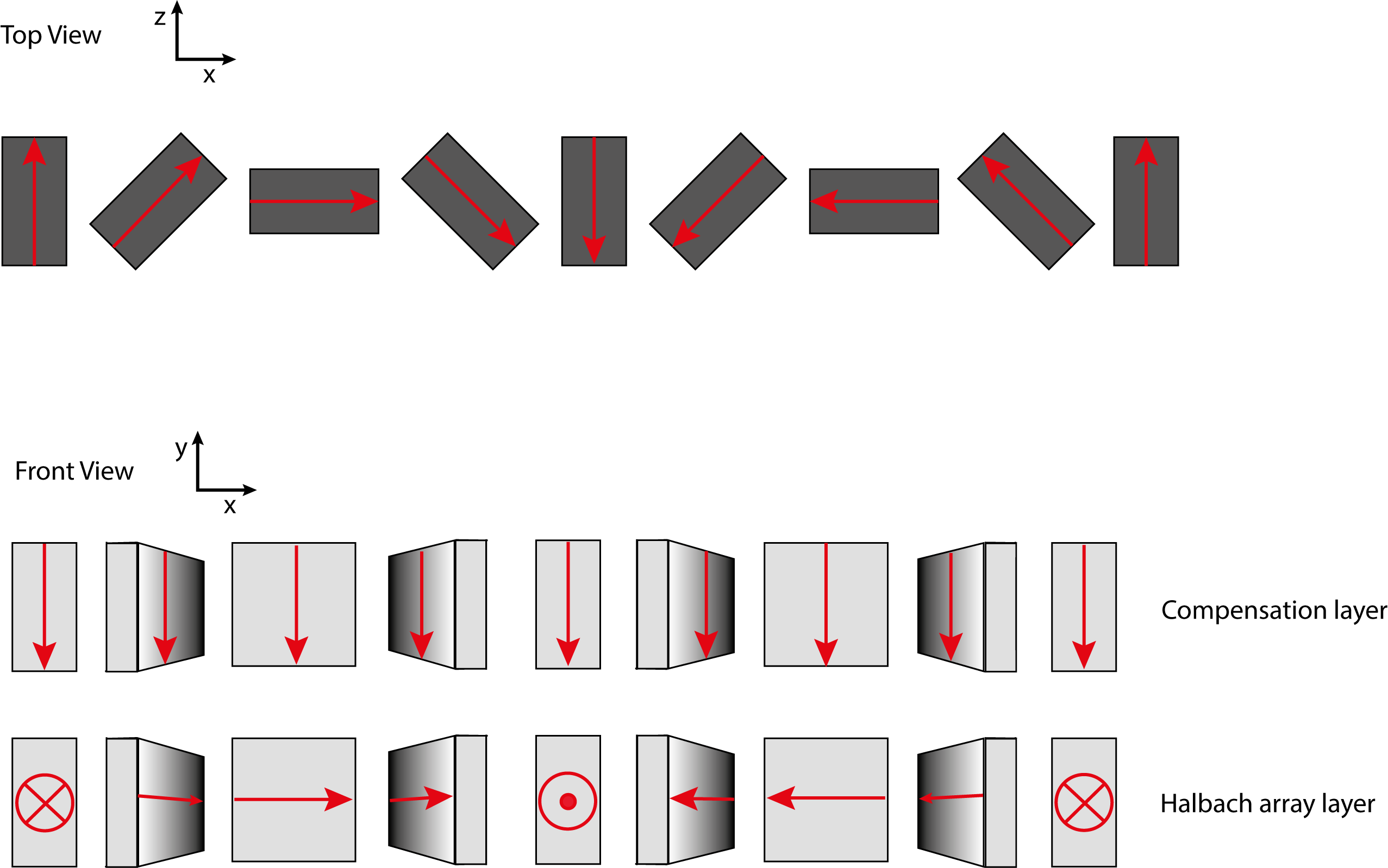}
	\caption{Diagram demonstrating the geometry of the dual-layer magnet configuration. The lower layer is comprised of the Halbach array, and the upper layer is a geometrically identical magnet array aligned in the negative y-axis (North pole pointing vertically towards the Halbach array).  The red arrows indicate the direction of the magnetic flux domains, while the red crosses and dots indicate arrows aligned into and out of the page respectively.}
	\label{fig:BasicMagnetDiagram}
	\end{figure}

\begin{figure}[H]
	\centering
	\subfloat[\centering]{{\includegraphics[width=\linewidth]{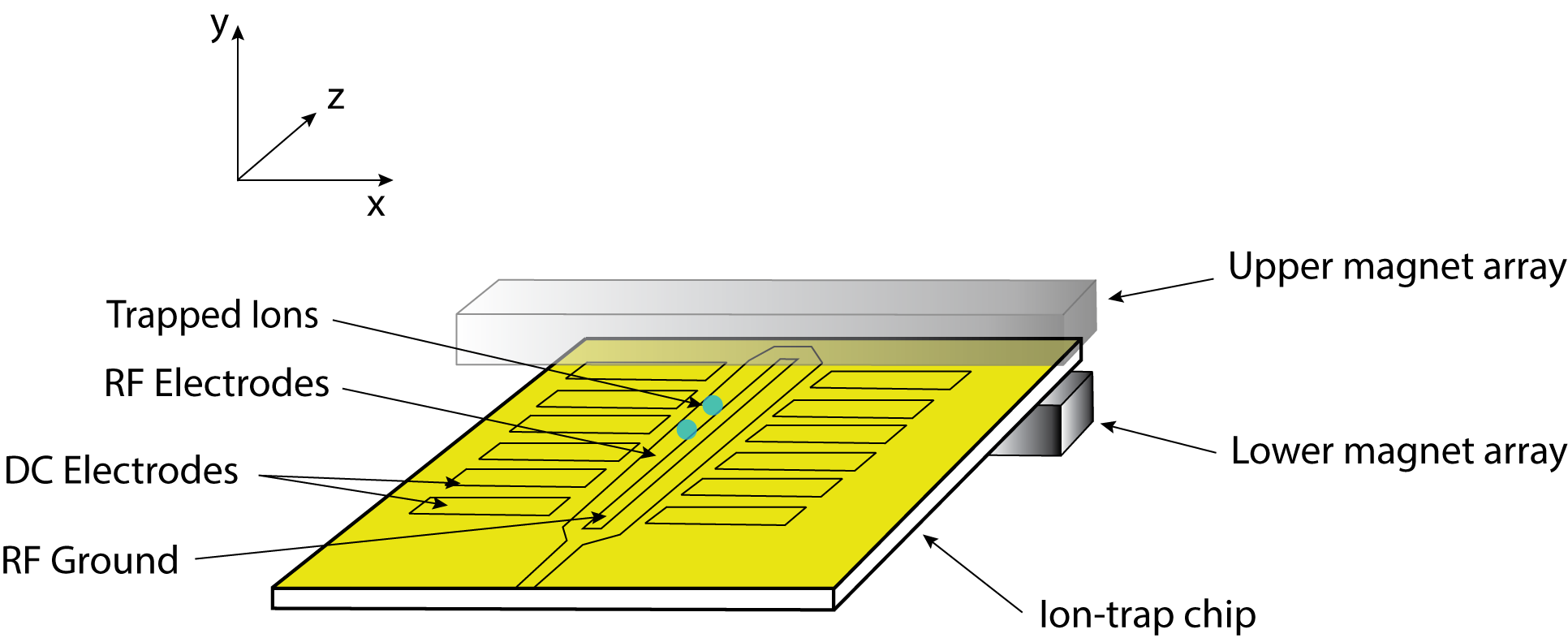}}}%
	\qquad
	\subfloat[\centering]{{\includegraphics[width=0.8\linewidth]{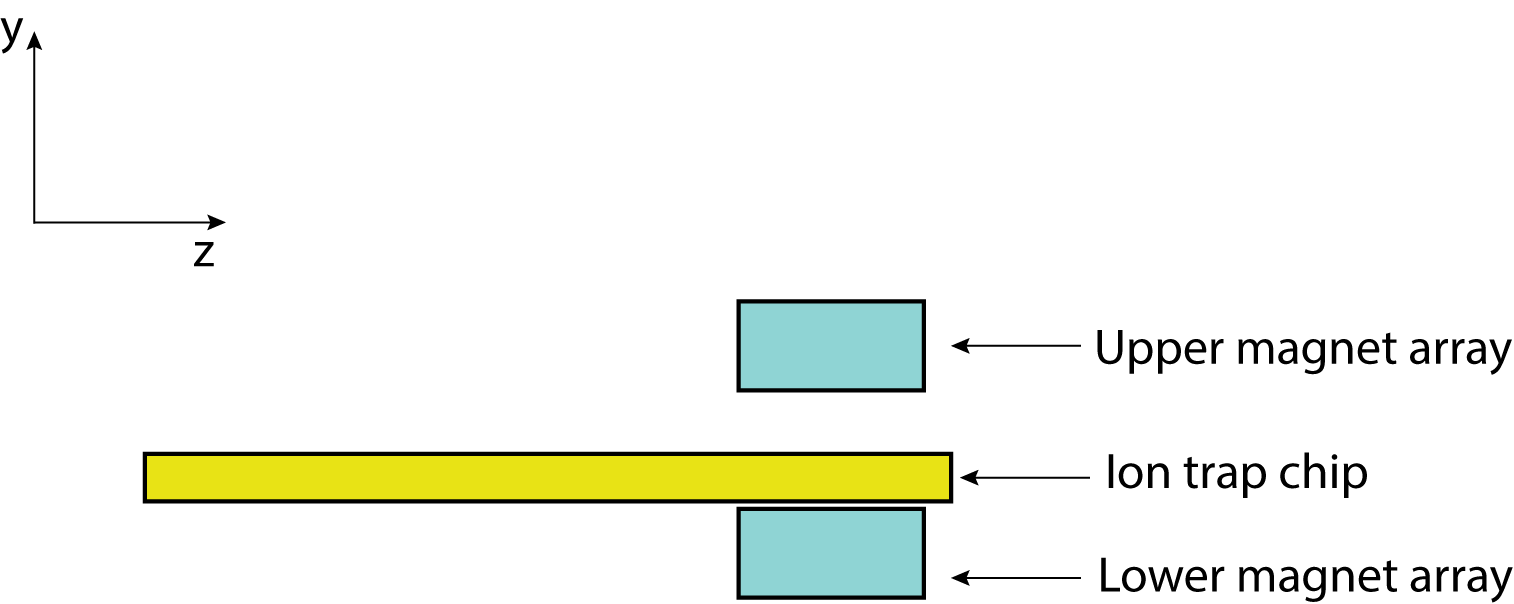}}}%
	\caption{Illustration of a linear, surface ion-trap orientation to magnet arrays showing principal axes for simulation and analysis. (a) Perspective view. (b) Side view.}
	\label{fig:trap_magnet}%
	\end{figure}

\begin{figure}[H]
		\centering
		\subfloat[\centering]{{\includegraphics[width=0.5\linewidth]{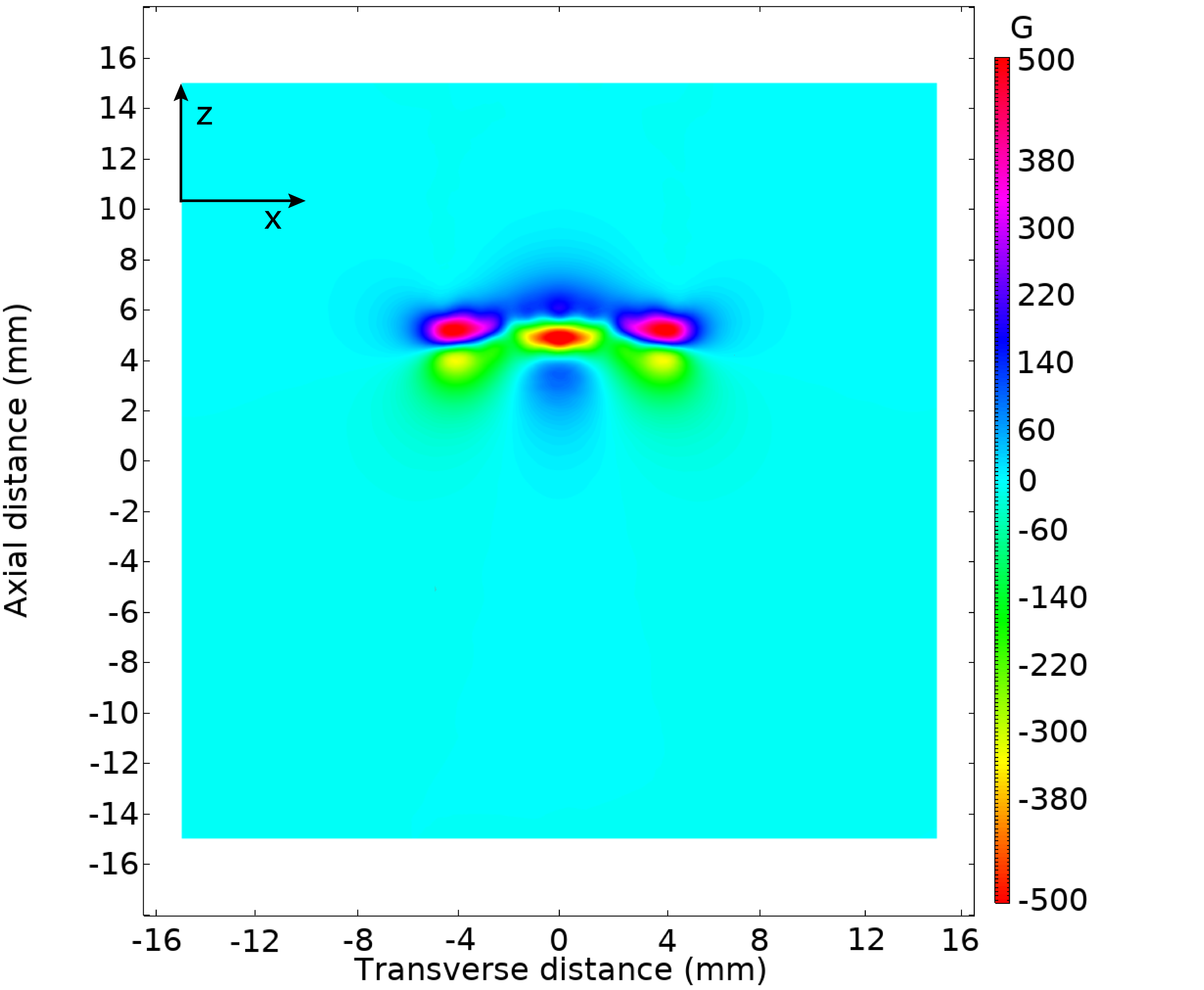}}}%
		\qquad
		\subfloat[\centering]{{\includegraphics[width=0.5\linewidth]{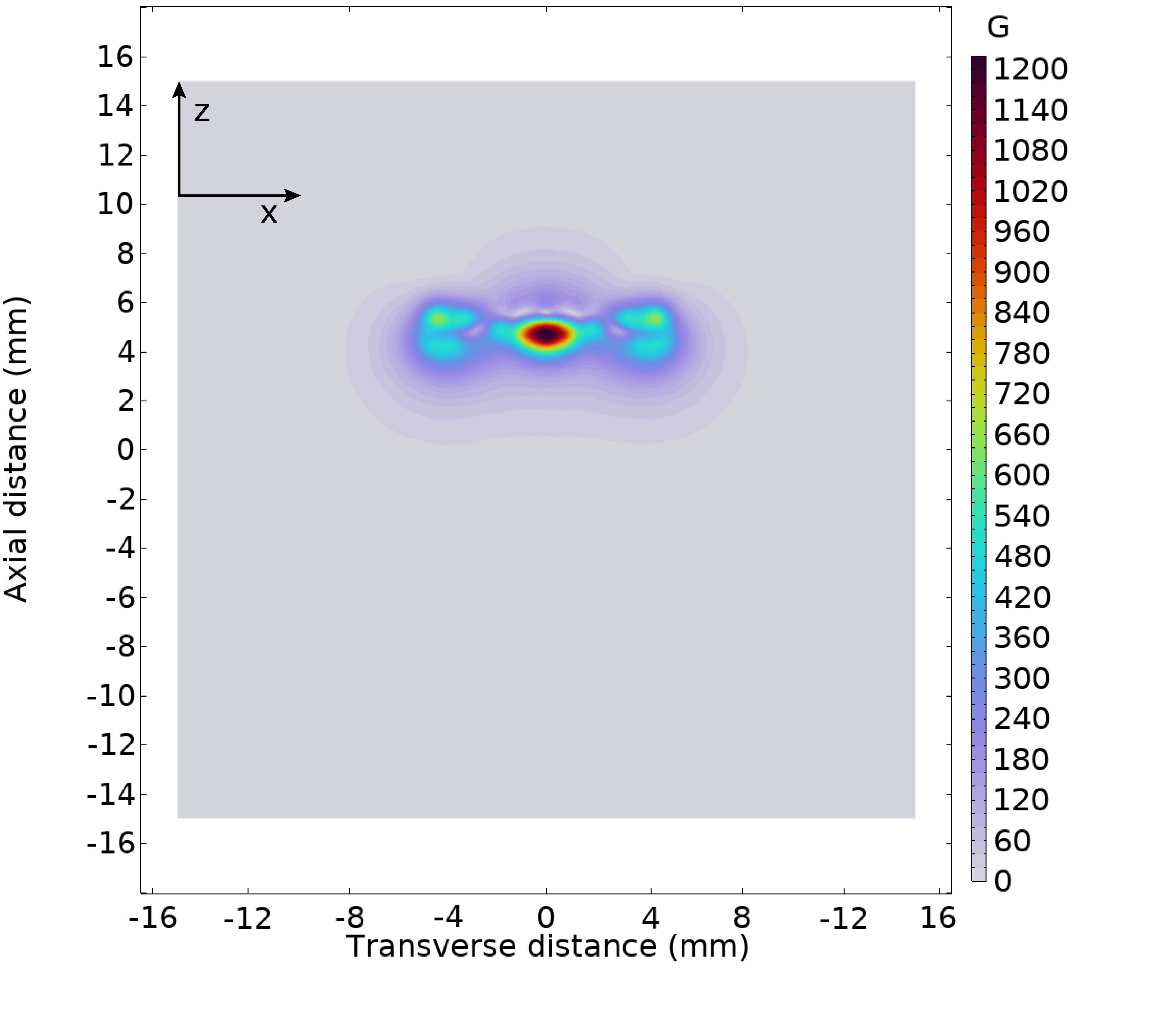}}}
		\caption{Contour plot showing the magnetic flux density, using the complete design, at a surface $0.5$ mm above the Halbach magnet array, representing a typical ion height plus the physical thickness of the surface trap. (a) Magnetic flux density in the axial direction (b) absolute magnitude of the magnetic field.  The color maps are given in units of Gauss.}
		\label{fig:ContoursMainDesign}%
	\end{figure}

\begin{figure}[H]
	\centering
	\subfloat[\centering]{{\includegraphics[width=0.45\linewidth]{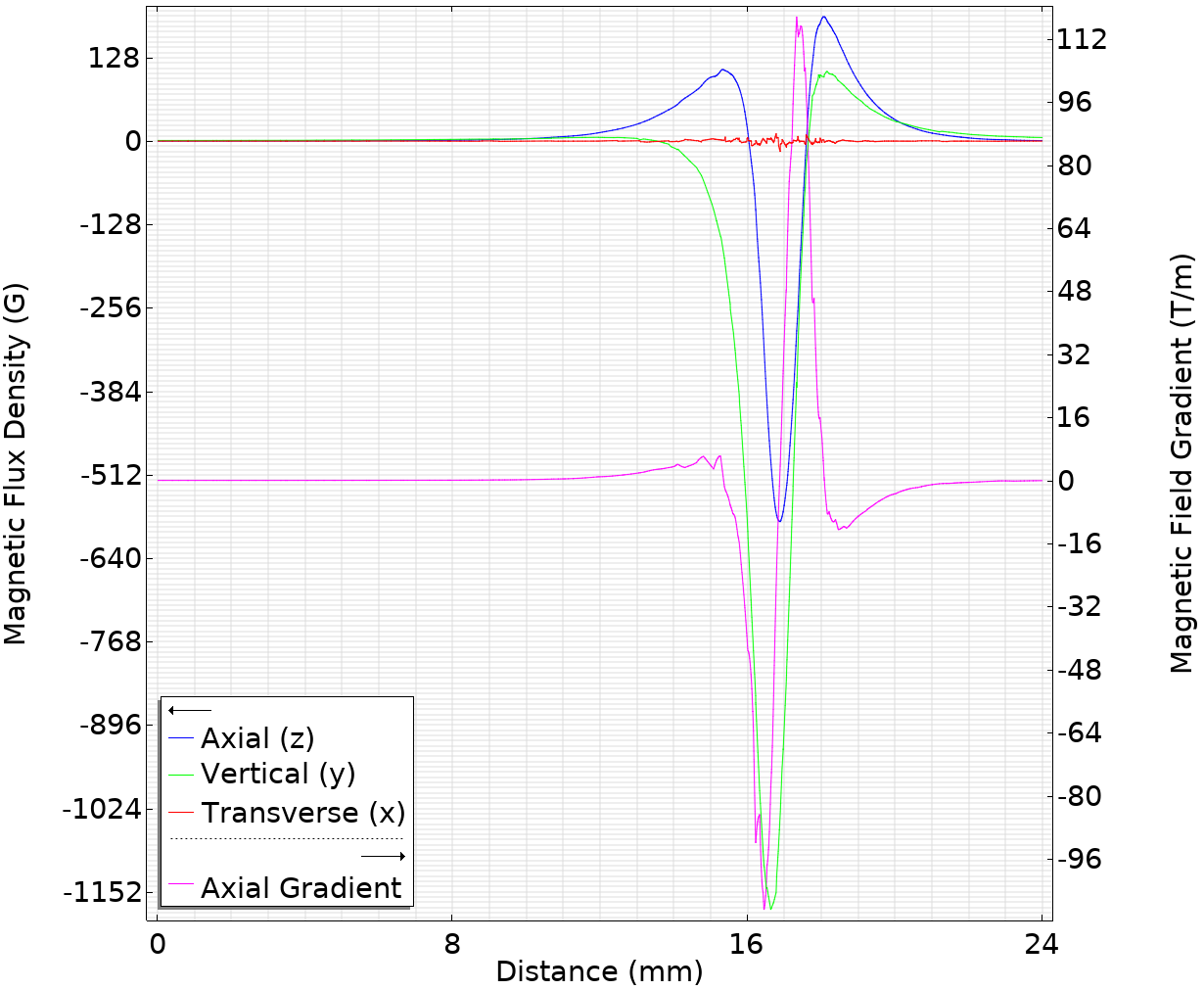} }}%
	\qquad
	\subfloat[\centering]{{\includegraphics[width=0.45\linewidth]{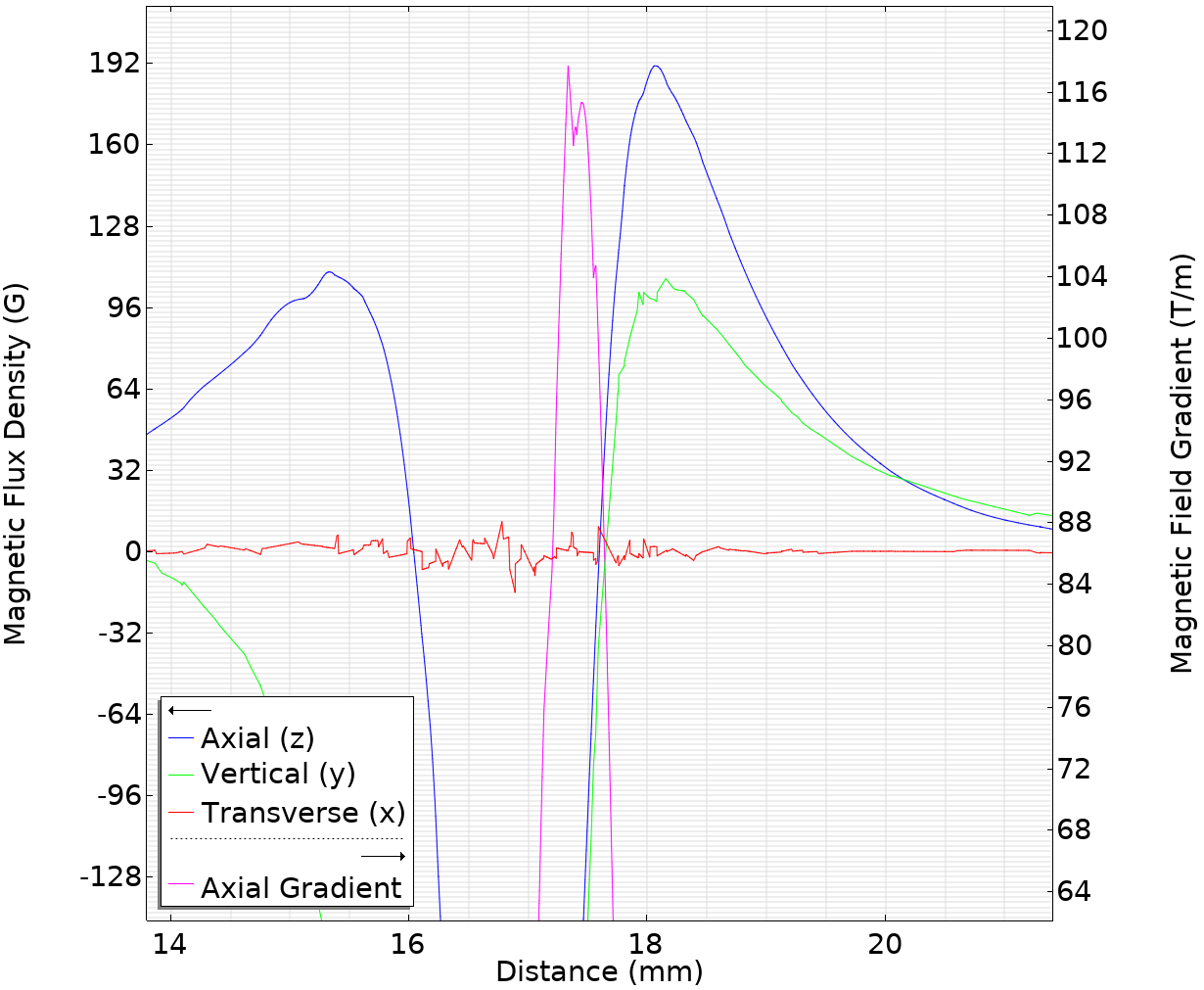} }}%
	\caption{ (a) Magnetic flux density contribution in the three axes: axial (z), vertical (y) and transverse (x), marked in blue, green and red respectively with magnetic field gradient in the axial direction (z) overlaid (magenta) showing the low field edge on the right side.  The intersection of the plane of the magnet edge is at $16$ mm in the simulation space.  (b) Zoomed region of interest where at $\approx 17.6$ mm the magnetic field has an effective null in three axes with a strong axial gradient, $1.6$ mm from the aligned edge of the magnet arrays.}
	\label{fig:HalbachDataPlots}%
\end{figure}

	The design features a Halbach array of nine segments, each consisting of a cuboid magnet of dimensions $0.5$ x $1$ x $1$ mm, and remanent magnetic flux density of $1$ T.  The magnetic domains are aligned, from parallel with the y-axis for the left-most magnet, as represented in the diagram, with $45^o$ rotations for each subsequent magnet in the array, to the right-most magnet, which is aligned parallel with the left-most magnet.  This array is placed in the x,z-plane, below the plane on which the field is analyzed, which is chosen arbitrarily, but reasonable for an ion height compatible with experimental parameters, in this case for mounting beneath a surface, micro-fabricated ion trap.  A compensation array of geometrically identical magnets is placed $2$ mm above the Halbach array but with the magnetic domains aligned in the negative y-axis (vertical) and with remanent magnetic flux density of $0.5$ T.  A diagram of the dual magnet array geometry is presented in figure \ref{fig:BasicMagnetDiagram}, and an illustration of the ion trap and magnet array arrangement is given in figure \ref{fig:trap_magnet} .  The contour plots and relevant gradient and magnetic flux density data for the field resulting from the simulation are displayed in figures \ref{fig:ContoursMainDesign} and \ref{fig:HalbachDataPlots} respectively.
	
	The magnetic field gradient in the axial direction is plotted as a function of axial position, in the plane of the arbitrarily designated ion height above magnets of $0.5$ mm.  The contributions of the magnetic flux density in the three principal axes were also plotted as a function of axial position at the ion height, shown in figure \ref{fig:ContoursMainDesign}.  The results show a small increase in the magnetic flux density in the axial and height axes, and a negligible contribution in the transverse axis.  The gradient increases sharply at the interface of the sparse and dense magnetic field regions above the edge of the magnets.  The vertical contribution (ion height direction), will still have a significant magnetic field offset which must be suppressed.  This offset can be compensated by a geometrically similar magnet array placed above the Halbach array, but with the remanent magnetic domains oriented co-linearly, and in the vertical direction.  This has the effect of significantly suppressing the vertical magnetic field offset while preserving the desirable properties of the Halbach array magnetic flux topography in the axial direction.

As presented in figure \ref{fig:HalbachDataPlots}, a magnetic field nil in all three axes can be achieved on the weak field side at the ion height.  Furthermore, this nil position is consistent with a large magnetic field gradient, nominally $\approx 90$ Tm$^{-1}$ from the simulations presented here.  Distinctly from the nil produced in the experimental setup previously discussed, here the nil can be reached by an ion moving through a region with a relatively small field offset in two axes and negligible field in the third.  Additionally, the magnetic flux density has a static profile with deterministic change over distance.  The magnetic flux density becomes negligible at distances on the order of $\approx 7$ mm from the magnet array.  This extinction of the magnetic field is a desirable quality when considering the scalability of the design in a shuttling QCCD architecture \cite{Lekitsch2015}.

While this design is promising, the undesirable magnetic field offset in the axial and vertical axes is still quite significant.  To make use of the desirable properties of this magnet configuration, the magnetic field offset in the plane of an ion being shuttled into and out of a gate region should be minimized such that use in the scalable quantum computing concept is viable.  Ideally, ions shuttled around a trap array in a scalable system should be exposed to a small, time-independent and homogeneous magnetic field during ion transport, to minimize qubit phase angle rotations $\phi$, for a qubit of the form $\Ket{0} + e^{i\phi}\Ket{1}$.  Homogeneity is desirable, as inhomogeneity in the field will also result in a time-varying field in the ion reference frame during ion transport through the field.  Unknown phase accumulation is a source of decoherence, dephasing error and subsequent limitation of $T_2$ coherence time \cite{Wineland1997}.  In practical systems, small inhomogeneity in the magnetic field will be unavoidable due to small systematic and random effects of the environment and components in the system e.g. from metal surfaces becoming slightly magnetized from exposure to the magnets, electrical noise from control electronics or sources in and around the laboratory.  The engineering tolerance of the magnetic field offset should be sufficient such that reasonable measures can be taken for compensation of offsets, and reliable tracking of the phase accumulation in the field.  This metric will vary dependent on implementation and design, but can be taken to mean compensable using a Helmholtz coil or other means of delivering a small magnetic field correction, which is feasible to implement experimentally.  
In this case, the compensable limit of the field will be taken as similar to that for the magnet configurations employed on previous experiments \cite{Peaks2023}\cite{Murgia2017}.  This will set the maximum compensable magnetic field offset of $60$ G, used as a specification for design optimization, and a useful threshold to consider a natural upgrade to an existing experimental design with minimal time and engineering cost.  It is worth noting however, that a scalable design or further experiment modification could make use of embedded solenoids on the ion trap chip or in-vacuum electromagnets for compensating the field, potentially allowing a greater degree of compensation, and lower power consumption/thermal load compared with using external compensation coils \cite{Siegele-Brown_2022}.  Such embedded coils could also be used to provide spatial homogeneity in the magnetic field in a desired region.  In this modified Halbach array paradigm, the change in magnetic field when shuttling into and out of the low field region should be minimized in order to minimize the cumulative effect of the magnetic field offset on the qubit phase during ion transport.  However, an advantage of using permanent magnets is that this change will be measurable, systematic and deterministic for each shuttling operation through the fixed magnetic field environment, due to the static nature of the remanent flux.  This means that the phase accumulation from shuttling through the region is deterministic and stable for a given ion shuttling rate, making it easily calculable after calibration measurements of the qubit phase for each region, even with manufacturing defects and local inhomogeneities between regions.  In the following section, an adjusted magnet geometry is considered in order to optimize for the gradient and magnetic field offset towards implementation in a proof-of-principle experiment, with focus on the best set of engineering compromises when considering a potentially scalable system.

In order to optimize the design, the geometry of the central magnets in the arrays were adjusted such that a compression in magnetic flux density could be achieved along the axial direction.  Simulating different geometries for the central magnet in each array then led to the optimized design presented in the following sections.

\section{Rhombic center design}
\label{sec:RhombicCentreGeometry}
\begin{figure}[H]
	\centering
	\includegraphics[width=\linewidth]{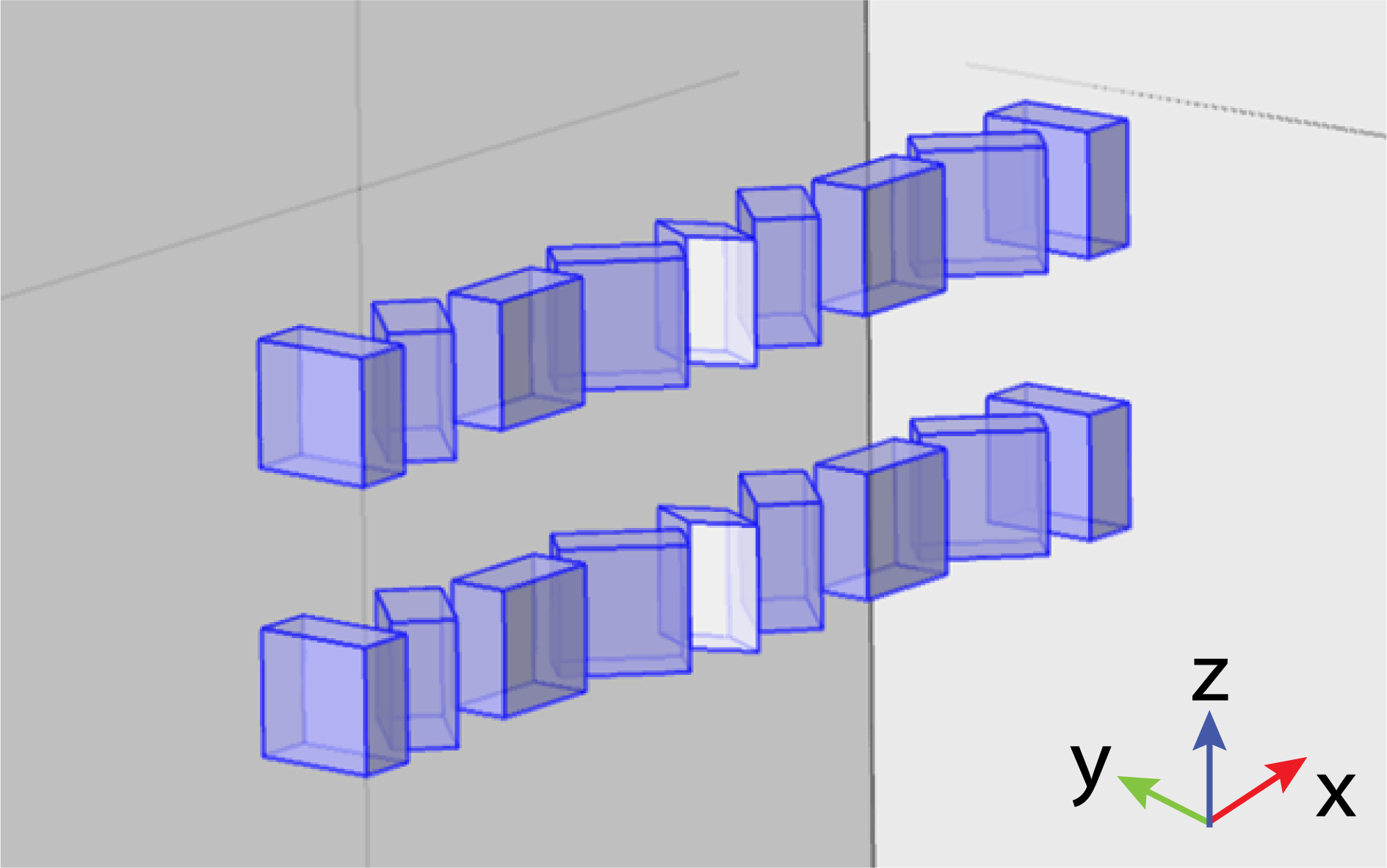}
	\caption{3D model of the magnet array configuration as presented in the COMSOL Multiphysics simulation package.  This adapted geometry features identical Rhombic prism for the central magnet in the two arrays with cross-section dimensions of $1$ mm x $0.5$ mm, and height $1$ mm.  The other magnets in the array maintain cuboid $0.5$ mm x $1$ mm x $1$ mm dimensions.}
	\label{fig:Rhombic3DModel}
\end{figure}

In the following section, a second magnet geometry with a similar dual array configuration is presented.  This design employs a modified magnet geometry from the na\"{i}ve array design to improve upon the desirable magnetic field characteristics.  Attention was given to maximizing the magnetic field gradient while minimizing the field offset in three dimensions.  Engineering compromises were also prioritized such that the most viable design was also realistic to fabricate and achieve for initial experiments, with a potential for scaling to larger quantum computing architectures.  Functionally, this means that the desired ion distance from the physical edge of the magnets must be suitable for laser beam access, with space allowed for mounting in a physical system with a surface ion trap.  The following design is similar in basic dimensions to the previous, but with the cuboid central magnet of both Halbach and compensation arrays replaced with a rhombic prism, with rhombic cross-section dimensions $1$ mm x $0.5$ mm, and height $1$ mm.  A 3D model of the magnet array geometry is presented in figure \ref{fig:Rhombic3DModel}.  The rhombic design of the central magnet in the array changes the magnetic flux profile around the key region at the edge of the array in the x,z-plane.

The rhombic central magnet alters the spatial characteristics of the magnetic field compared with the cuboid, the magnetic field offset at the approach to the weak field side, in the axial and vertical directions, is significantly reduced and yields a nil position in three axes at $18$ mm, but coincides with a reduction in the axial magnetic field gradient, $10$ Tm$^{-1}$ at the nil position.  While the reduced magnetic field offset is desirable, the axial gradient is not ideal, as the effective Lambe-Dicke parameter scales linearly with the gradient, and maximizing this parameter is desirable to increase spin-motion coupling, allowing multi-qubit interactions and increasing the fidelity in long-wavelength quantum gate control schemes \cite{Mintert2001, Lake2015}.  However, there is still a significant parameter space associated with this geometry which is open to optimization.  It is then desirable to optimize for a magnetic field nil in three dimensions, a suppressed magnetic field in the region where a shuttling operation would be effected, and an axial magnetic field gradient sufficient to provide the required spin-motion coupling for high-fidelity gates at the field nil.


\begin{figure}[H]
	\centering
	\subfloat[\centering]{{\includegraphics[width=0.45\linewidth]{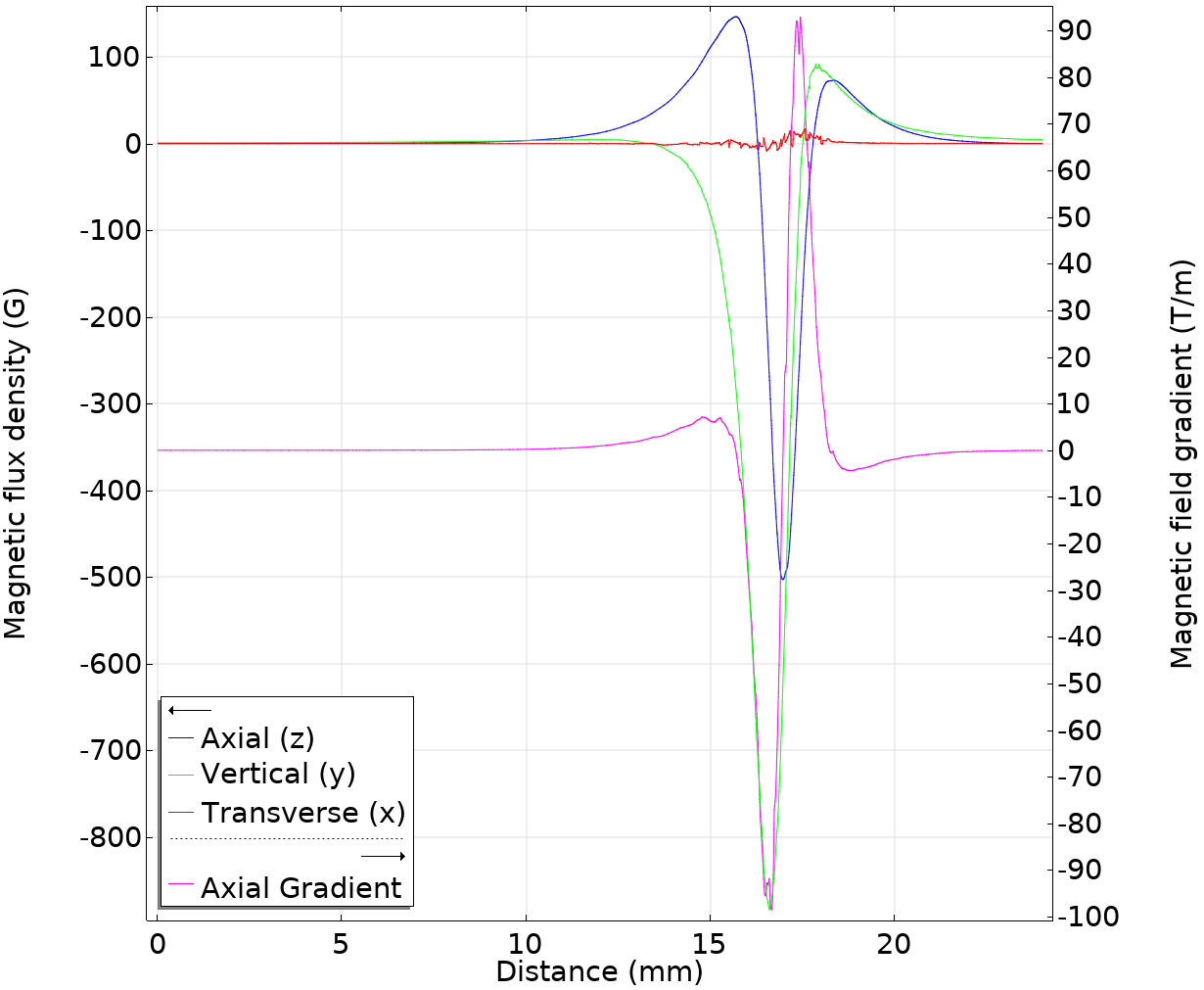} }}%
	\qquad
	\subfloat[\centering]{{\includegraphics[width=0.45\linewidth]{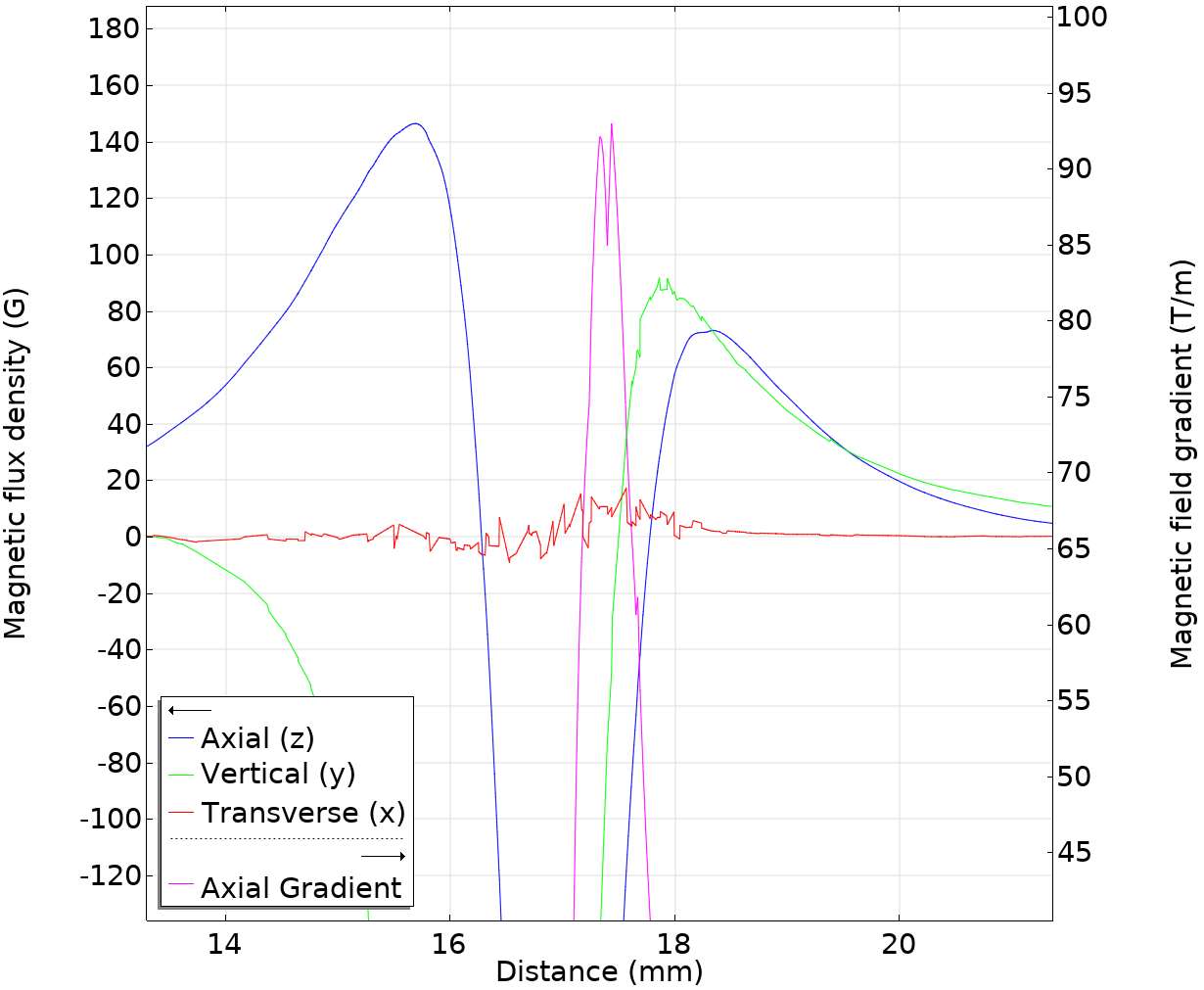} }}%
	\caption{ (a) Magnetic field gradient in the axial direction (y) for the rhombic prism center magnet Halbach array design, plotted with magnetic flux density contribution in the three axes: axial (y), vertical (z) and transverse (x), marked in blue, green and red respectively.  The intersection of the plane of the front magnet edge is at $16$ mm.  (b) Zoomed region of interest showing the gradient approaching the array on the weak field side.}
	\label{fig:RhombicUnoptimisedPlots}%
\end{figure}

	\begin{subsection}{Parameter optimization}
	\label{sec:HalbachParameterOptimisation}
	
	The global and relative positions of the arrays, the remanent flux of each magnet and the specific geometry of magnets in the array are but a few examples of potentially adaptable parameters.  Starting from the very promising modified Halbach array with rhombic central magnets presented previously, a process of manually sweeping the parameter space for each of the geometric and physical constraints was performed.  Extrema for the desired properties were sought by iterative changes in the simulation parameters.  In particular, the magnetic field offset in three dimensions should be minimized for the active side of the array i.e. the position to which the ion should be moved into and out of in a scalable system.  The magnetic field gradient at a magnetic field nil or useful ion position should also be sufficient to provide a comparable effective Lamb-Dicke parameter for high-fidelity quantum control using the state-dependent coupling scheme \cite{Mintert2001}.  The parametric constraints for the optimization process are discussed in following.
	
	\begin{subsubsection}{Separation of magnet arrays in the vertical axis}
		\label{sec:ParamMagnetSeparation}
		
		The separation of the two Halbach arrays in the vertical axis is a natural starting point for optimization of the field as this is the direction in which the field requires compensation compared with traditional dipole and quadrupole magnet setups \cite{Lake2015, Kawai2016}.  To simplify exploration of the parameter space, the lower Halbach array is taken to be fixed, and the ion height also fixed at $0.5$ mm in relation to the plane tangent to the top surface of the lower array, referred to henceforth as the ``base-plane''.  This gives a reasonable starting point and bounds the search such that a trap of similar design to that used in reference \cite{Peaks2023} can feasibly be fixed and operated in a mounting structure relative to the positions of the Halbach arrays.  The position of the upper Halbach array is then simulated at varying distances above the base-plane, while the relative position in the other two axes (transverse, x, and axial, y) are fixed in alignment with the lower Halbach array.  The geometry is illustrated in figure \ref{fig:VertAxMagOptimise}.  The range over which this distance can be varied has a lower bound set by the surface of an ion trap fixed atop the base-plane.  It is also possible to consider the ion height a lower bound here, however, it is shown in the previous analysis in section \ref{sec:GeneralDesign}, that it is possible for the desirable ion position to be separated from the magnet arrays by several millimeters in the axial direction, thus not limiting the upper array position to be above the ion height and allowing a greater range of positions to be explored.  The upper limit is only bounded by physical practicalities of mounting and the space inside the vacuum chamber for a hypothetical experimental setup.  However, the interaction of the magnetic field contribution from this array diminishes with distance, to the ion, which constrains the functional parameter space considerably.  The separation of the two arrays was chosen to be $2.25$ mm after a parameter sweep, constrained by minimizing the magnetic field offset in three dimensions, and maximizing the gradient with the other dimensions fixed.
		\vskip4mm
		\begin{figure}[H]
			\centering
			\includegraphics[width=\linewidth]{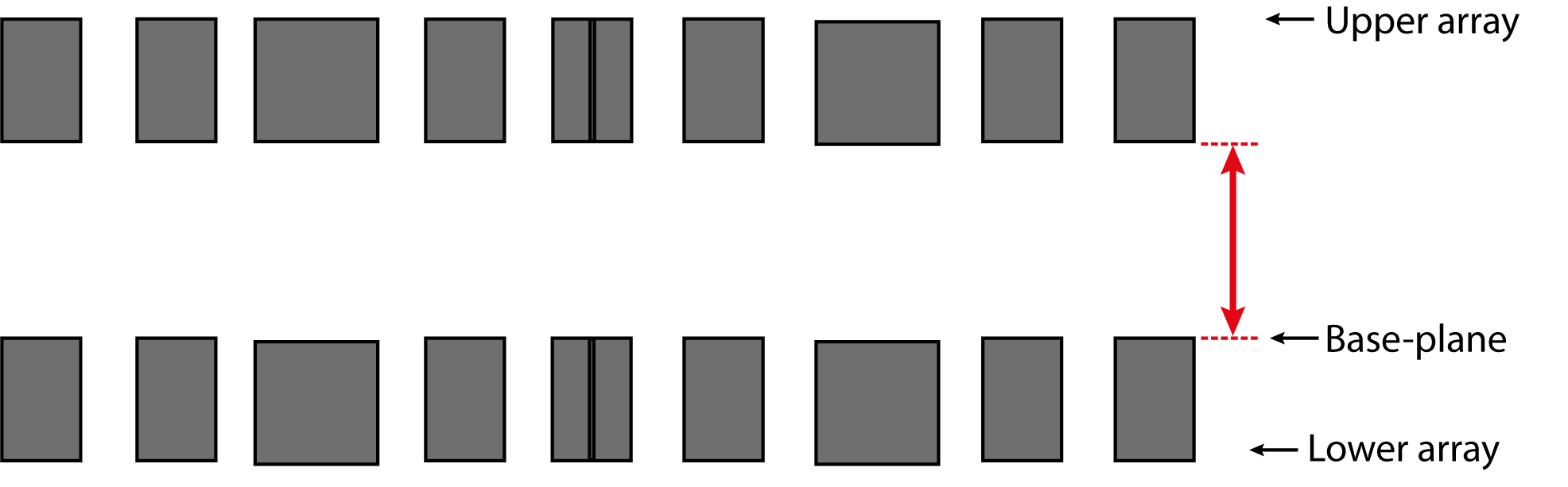}
			\caption{Illustration showing the parameter to vary for optimization of the magnet array separation in the vertical axis (red arrow).  The base-plane is indicated for clarity.}
			\label{fig:VertAxMagOptimise}
		\end{figure}
	\end{subsubsection}
	
	\begin{subsubsection}{Relative position of magnet arrays in the axial direction}
		\label{sec:ParamMagnetRelativeAxial}
		The relative position of the two arrays can also be adjusted in the axial, y, direction.  Once again, the base-plane and lower Halbach array were fixed in position and the upper array moved relative to this reference for simplicity and consistency of simulation.  The starting position was taken to be complete alignment with the lower Halbach array in the axial and transverse axes, with the offset in the vertical direction fixed, initially at an arbitrary starting position of $2.5$ mm, and later at the best separation parameter identified, at $2.25$ mm.  Small adjustments to this parameter shift the relative position of equal field magnitude in axial, y, and vertical, z, directions.  This was, therefore, a useful parameter to fine tune the magnetic field in three axes for the optimal magnetic field nil position, where the ion would be held in an experiment or scalable quantum computing system.  After scanning this parameter while fixing the others for each simulation with a range of setups, the optimal, global configuration identified with the other parameters was found to have no offset in the relative axial positions of the arrays.  This was likely because it proved practically easier to modify the relative vertical position to achieve a similar effect when manually fine tuning the simulated parameters.
		\vskip4mm
		\begin{figure}[H]
			\centering
			\includegraphics[width=0.7\linewidth]{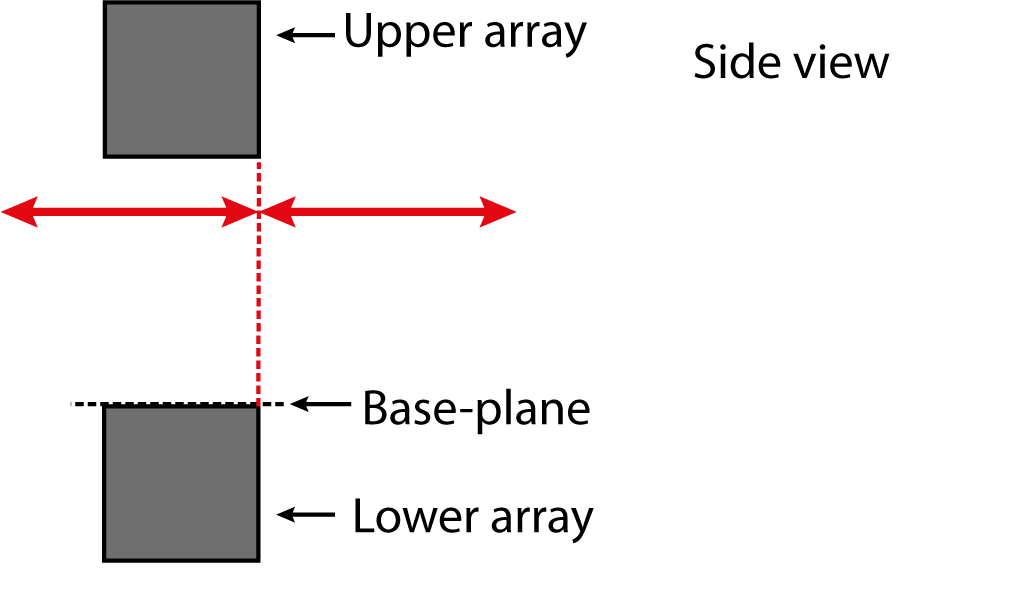}
			\caption{Illustration showing the parameter to vary for optimization of the relative position of magnet arrays in the axial direction (red arrow).  The base-plane is indicated for clarity.}
			\label{fig:AxialAxMagOptimise}
		\end{figure}
	\end{subsubsection}
	
	\begin{subsubsection}{Transverse separation of magnets within arrays}
		\label{sec:ParamMagnetArrayTransverseSeparation}
		Within each of the Halbach arrays, relative separation between each magnet can be adjusted.  For a typical Halbach array configuration, it is normal to set equidistant spacing between each magnet in the array.  The spacing that is set between magnets changes the overall size of the near-field region, taken here to be a region of magnetic flux density much greater than a typical background magnetic field in the lab, $\approx 0.1$ G, thus $|B| \gg 0.1$ G.  Given the desired properties of the modified Halbach array for a scalable system, an ideal scenario would be negligible magnetic field and field gradient anywhere except for the specific region in which the quantum control would be applied to the ion.  Additionally, the magnetic field gradient which can be achieved, for any given remanent magnetic flux density, is proportional to the ratio of the dimensions of the field source to the distance from the source.  For these reasons, the Halbach arrays are compressed in this dimension as much as possible.  A physical lower limit is then imposed on the magnet size and separation by physical manufacturing tolerances for permanent magnets of this size and shape, and an achievable mounting structure for a demonstrable experimental setup.  The mounting structure may require machining in titanium and/or tungsten for the desirable magnetic and thermal properties (discussed in section \ref{sec:MountingStructure}).  Due to the limited, potentially artisanal nature of much of the required build process for an initial implementation experiment, the transverse separation of the magnets was left with some margin for error in manufacturing.  The center of each magnet is thus separated by $1.5$ mm from the center of each adjacent magnet in both upper and lower Halbach arrays.  It is noteworthy that improved manufacturing tolerances might be achieved with industrial fabrication techniques, and thus further optimization of the magnetic field properties.
		\vskip4mm
		\begin{figure}[H]
			\centering
			\includegraphics[width=\linewidth]{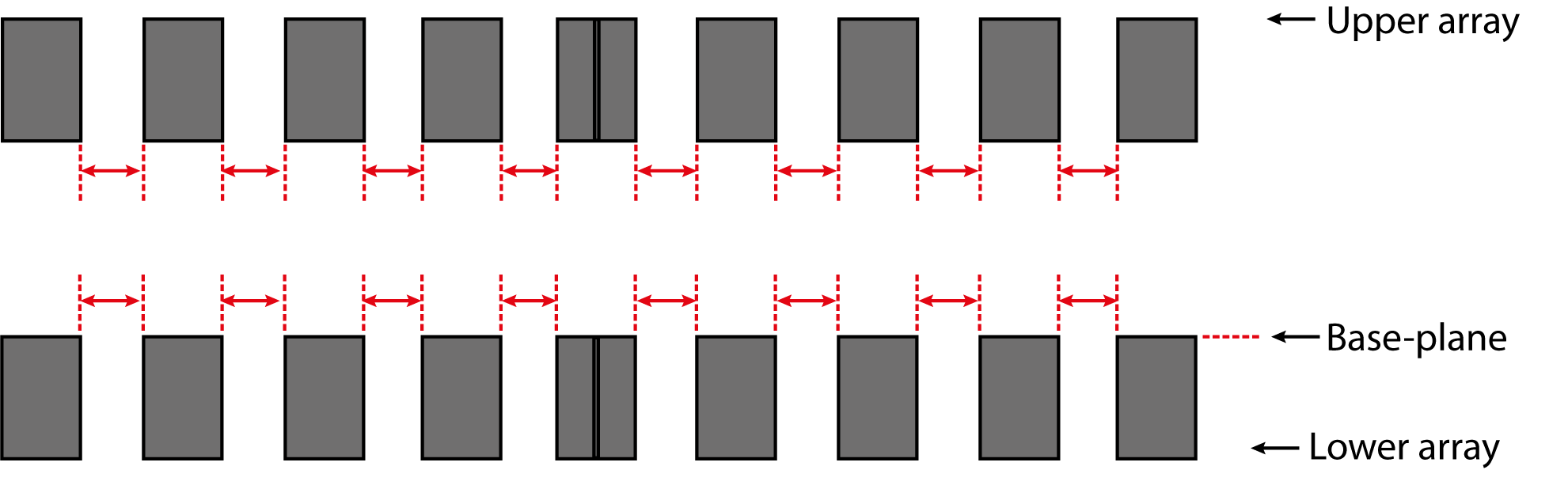}
			\caption{Illustration showing the parameter to vary for optimization of the magnet separation within arrays (red arrow).  The base-plane is indicated for clarity.}
			\label{fig:TransverseMagOptimise}
		\end{figure}
	\end{subsubsection}
	
	\begin{subsubsection}{Remanent magnetic field of magnets in array}
		\label{sec:ParamRemanentMagneticFlux}
		The remanent magnetic field of the magnets in the array has a linear proportional effect on the local magnetic field strength at the ion position.  The field gradient is linearly dependent on the remanent magnetic field but quadratically dependent on the distance of the ion position to the source.  It is thus more beneficial to focus on reducing the size of the magnet arrays and the ion proximity to the array than to maximize the strength of the magnets.  Furthermore, to compensate for the offset in magnetic field, effectively, the remanent flux density of the magnets in the upper and lower arrays should be matched as closely as possible.  For the purpose of these simulations, the remanent magnetic field of each magnet was taken to be $1$ T, as it was the maximum value available from custom Samarium-Cobalt (SmCo) magnet suppliers, cut to the specific geometries desired for this design.  
	\end{subsubsection}
	
	\begin{subsubsection}{Other parameters}
		\label{sec:ParamOtherConsiderations}
		All other parameters were fixed during this optimization process in order to reasonably bound the vast space of possible configurations, both for the purpose of practicality and to focus on the parameters which appeared to be the most prevalent in terms of desirable corrections to the relevant system properties.  This includes adjustment of relative rotation angle and position of the two arrays about the vertical axis, or asymmetric offsets in the transverse direction, translations and rotations in this plane collapse the symmetry of the system, required to the provide the desired cancellation of the field.  These geometric transformations also increase the footprint of the device which is undesirable from the experimental design and scalability considerations previously discussed.  Rotations of each Halbach array in the transverse direction were also not applied as the field nil is desired at a location simple to implement with a surface ion trap and magnet mounting structure.  In addition, this parameter is invariant to having magnets with a remanent magnetic field with the poles aligned in a different orientation relative to the shape/geometry of each magnet, which was a simpler solution given options available from suppliers.
	\end{subsubsection}
	
	\begin{subsubsection}{Result of parameter optimizations}
		The modified Halbach array field was simulated iteratively, with adjustments to the optimizable parameters in each case.  Initially, larger increments for course adjustment within the total parameter space were applied, followed by small adjustments to fine tune the desired field characteristics to a minimal, three dimensional, magnetic field nil position and a strong axial gradient.  Plots of the magnetic field gradient, and offset in three dimensions, resulting from a final simulation of the optimized parameter geometry are presented in figure \ref{fig:RhombicPlots}.
		
		\begin{figure}[H]
			\centering
			\includegraphics[width=\linewidth]{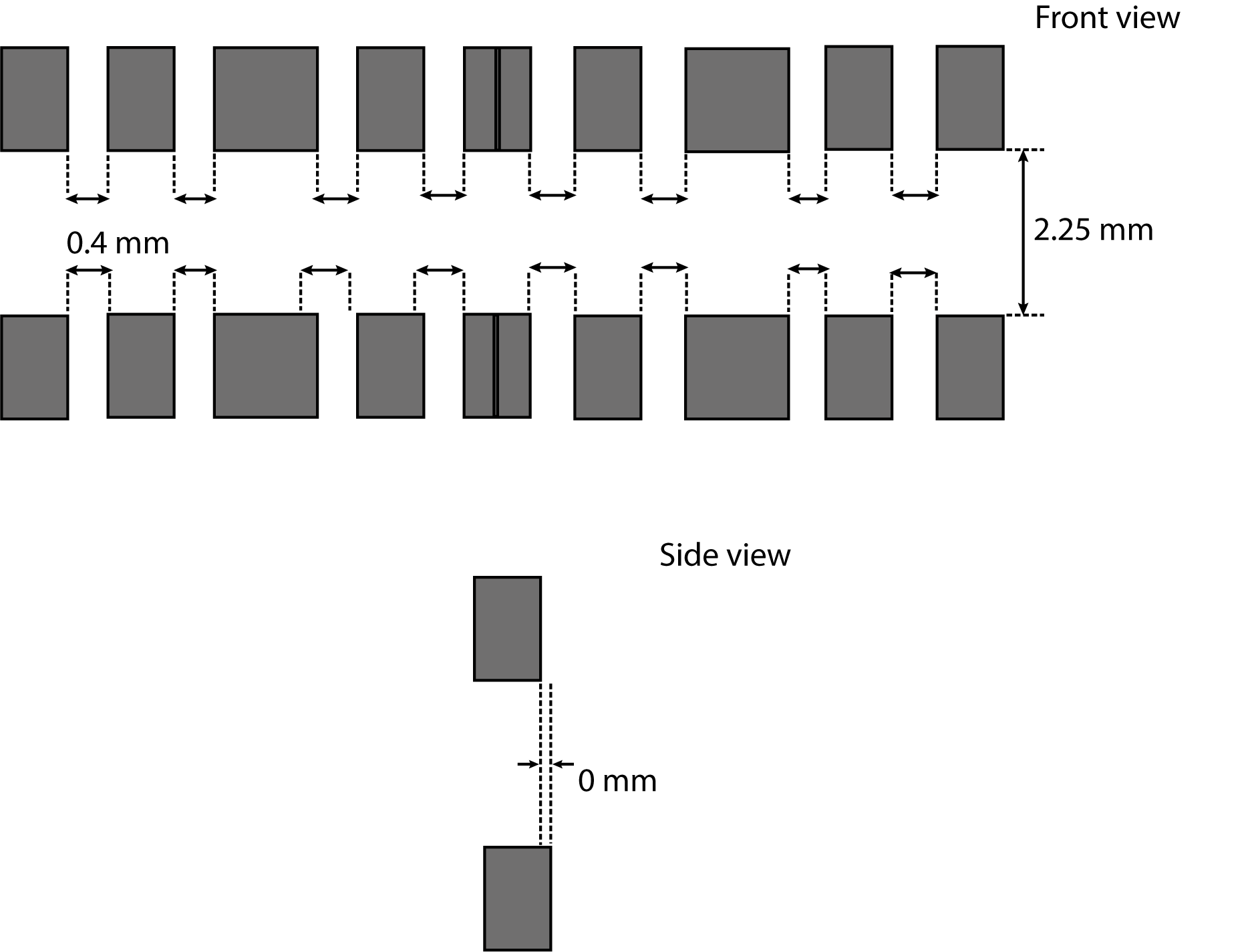}
			\caption{Illustration of the modified Halbach array dimensions with geometry optimizations providing relevant measurements.}
			\label{fig:OptimisedGeometryMeasurements}
		\end{figure}
		
		\begin{figure}[H]
			\centering
			\subfloat[\centering]{{\includegraphics[width=0.45\linewidth]{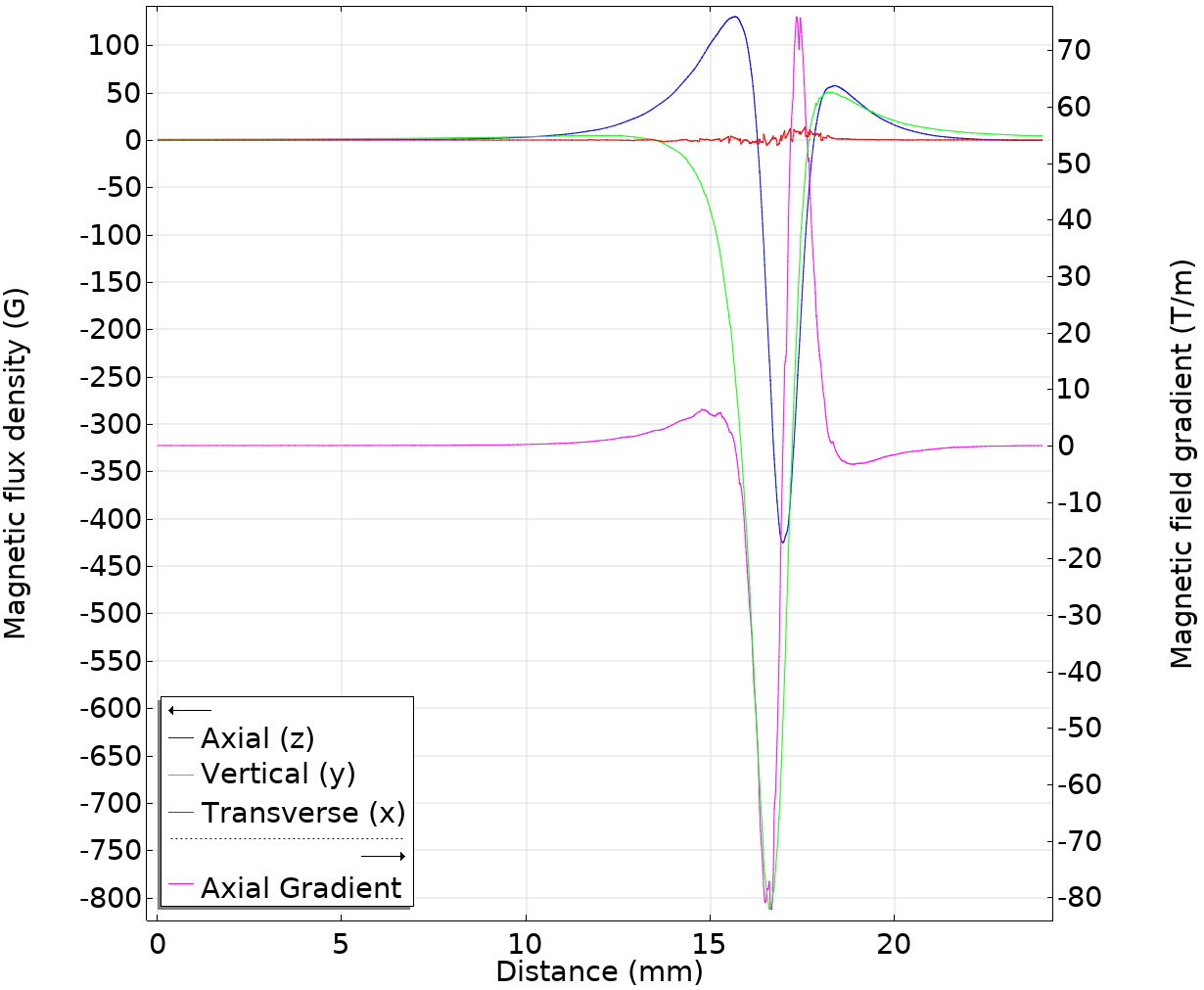} }}%
			\qquad
			\subfloat[\centering]{{\includegraphics[width=0.45\linewidth]{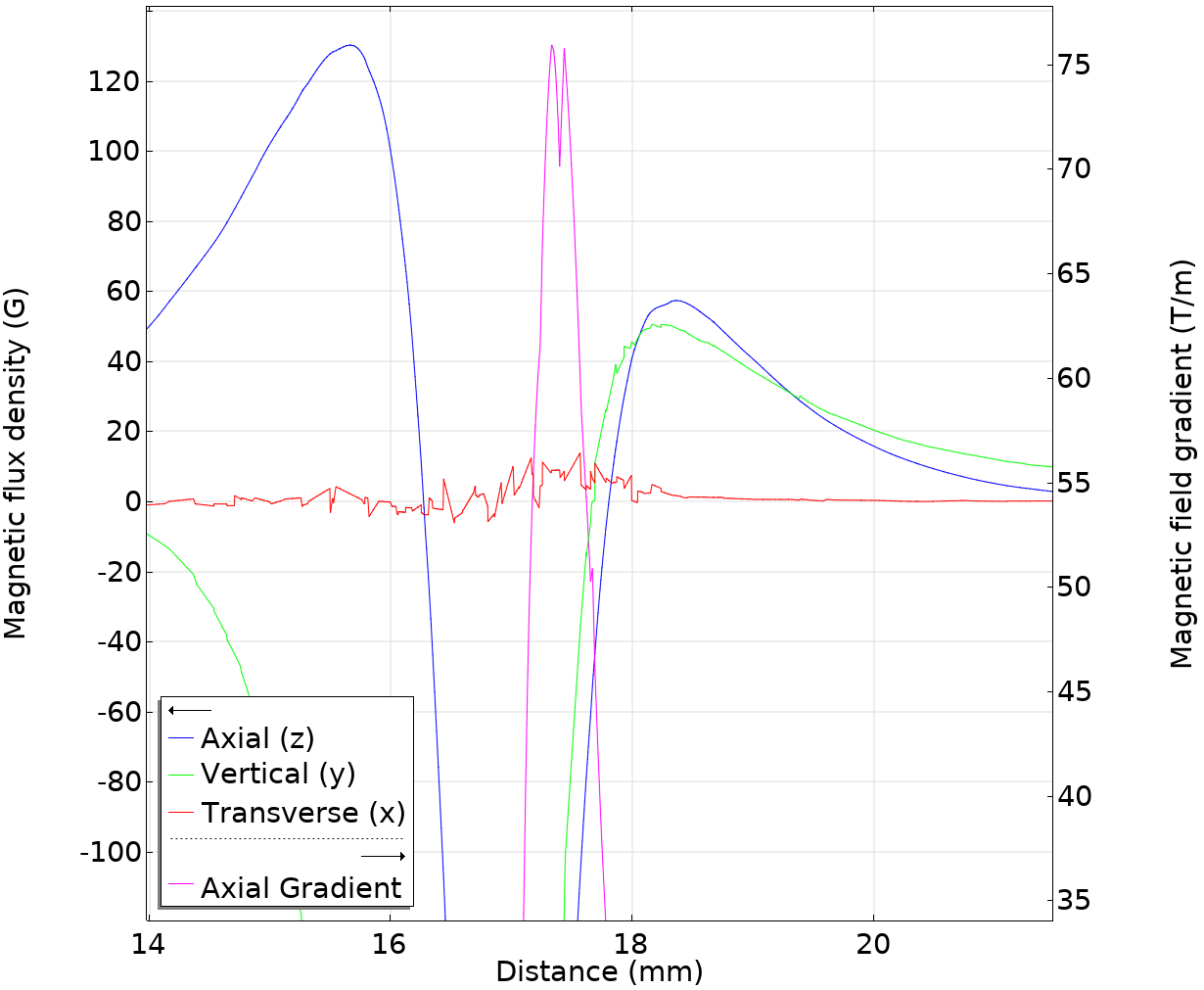} }}%
			\caption{ (a)  (a) Magnetic field gradient in the axial direction (y) for the modified version of the rhombic center array with parameter optimizations, plotted with magnetic flux density contribution in the three axes: axial (y), vertical (z) and transverse (x), marked in blue, green and red respectively.  The intersection of the plane of the front magnet edge is at $16$ mm.  (b) Zoomed region of interest showing the gradient approaching the array on the weak field side.}
			\label{fig:RhombicPlots}%
		\end{figure}
		
		The magnetic field gradient at $z = 17.6$ mm, $1.6$ mm from the magnet edge in the axial direction, a magnetic field gradient of $51$ Tm$^{-1}$ at a negligible magnetic field in three principal axes.  The approach to this point features a modest increase to $\approx 50$ G and $55$ G in the vertical and axial directions respectively.  This magnetic field offset is now sufficiently low that conventional compensation techniques such as magnetic field coils near the device can compensate for the offset.  Furthermore, the field changes with a well-defined character at distances throughout the setup.  Once again, the magnetic field offset becomes negligible at distances on the order of $\approx 7$ mm from the magnet array.  For the final design, both Halbach arrays were composed of the same number, size and shape as those shown using the rhombic center geometry above.  Figure \ref{fig:Rhombic3DModel}, shows the basic shape and configuration of the magnets.  The cuboid magnets are $0.5$ mm x $1$ mm x $1$ mm, width, length and height respectively.  The central, rhombic prism magnet in each array measures $0.5$ mm on each diagonal edge of the rhombic shape cross-section, and $1.0$ mm in height.  The remanent magnetic flux density of each magnet in the lower array is set at $1$ T, with magnetic domains aligned parallel with the y-axis (axial) for the left-most magnet, and the same for each magnet in the array relative to geometry, with $45^o$ rotations applied for each subsequent magnet.  The upper array has all magnetic domains aligned with the north pole pointing anti-normal with the base-plane, and with a remanent magnetic flux density of $0.5$ T.  The separation between the base-plane and the plane, on which the bottom surface of the upper array lies, is $2.25$ mm, providing ample space for an ion trap and mounting, discussed in section \ref{sec:MountingStructure}.  Both arrays are aligned flat and parallel with the base-plane, as in previous iterations.  For clarity, the geometry and measurements are given in figure \ref{fig:OptimisedGeometryMeasurements}.
	\end{subsubsection}
\end{subsection}

\section{Mounting Structure}
\label{sec:MountingStructure}
\begin{figure}[H]
	\centering
	\includegraphics[width=\linewidth]{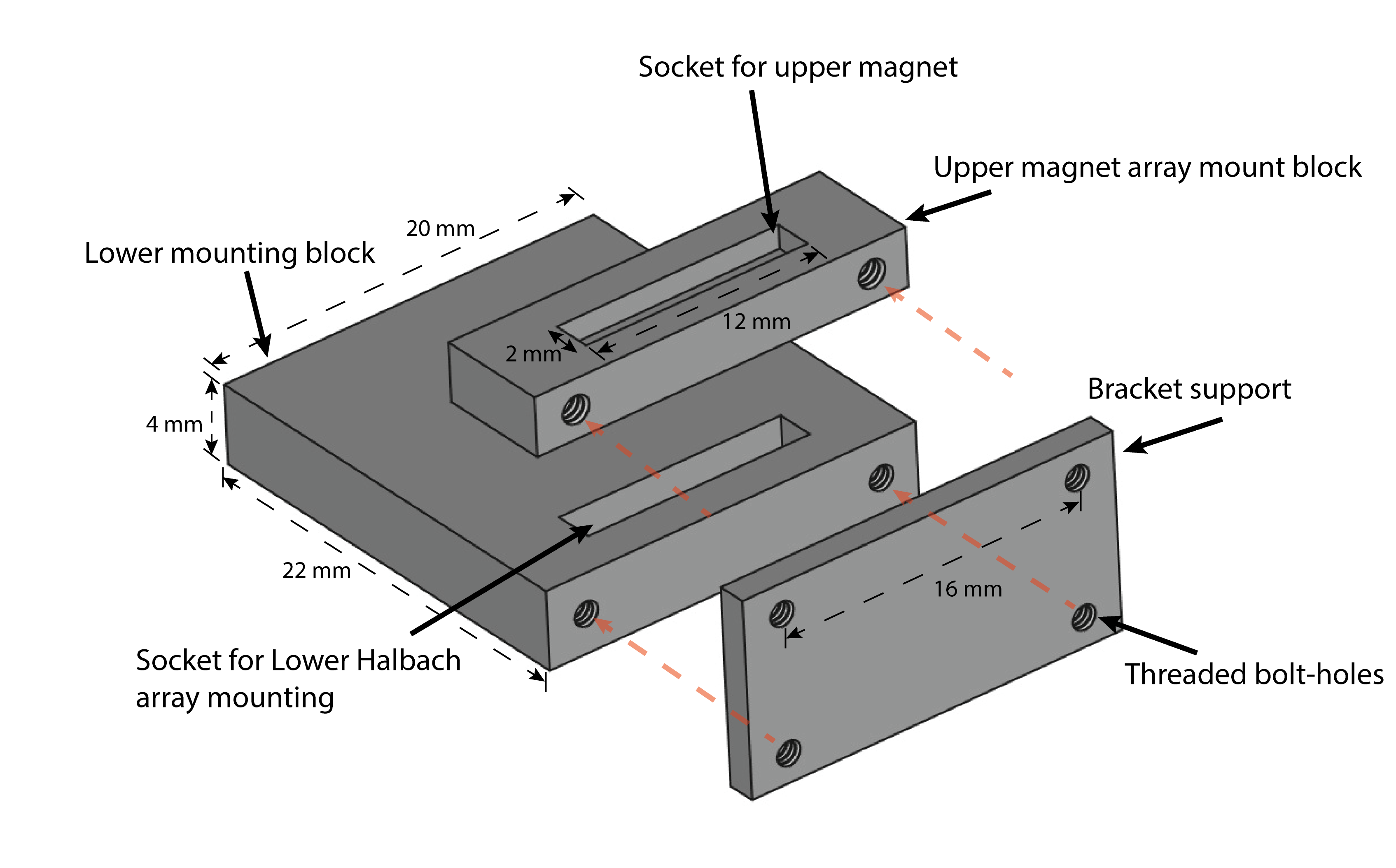}
	\caption{CAD model of the three-piece mounting structure for the Halbach array magnet design.  The lower Halbach array is made out of tungsten, whereas the other two parts would be machined from titanium for ease of manufacturing and ensuring tolerances are met for the desired alignment.  The red arrows indicate the the mating of the parts and position of bolts for assembly.}
	\label{fig:HalbachMounting}
\end{figure}

A concept design for the magnet and ion trap mounting is given in figure \ref{fig:HalbachMounting}.  The mounting structure for the magnet arrays and ion trap is composed of three parts described in the following section.

\begin{subsection}{Lower Halbach array and trap mounting block}
	The lower mounting structure could be made of tungsten or titanium dependent upon the application, cut to shape by Electro-Discharge Machining (EDM) wire-erosion or standard machining respectively. Tungsten is a desirable material both for low magnetic susceptibility and small coefficient of thermal contraction when applied in an experiment where the surface ion trap is cooled cryogenically.  Titanium is much easier to cut and machine while having somewhat less favorable magnetic and thermal characteristics for the application, thus is a reasonable second choice for parts requiring finer tolerances or more complicated shapes.  The Halbach array would be secured in the mounting block either by application of a single-part epoxy, or by bonding with an indium solder.  Both bonding techniques have merits when considering the temperature range and low pressure application for the experiment, able to withstand the temperature of the bake procedure to achieve UHV, $\gtrapprox 415$ K, and temperatures down to $\approx 40$ K to operate the experiment cryogenically.  The bonding must also be UHV compatible such that out-gassing will not continue to occur, compromising the required vacuum pressure.  A microfabricated, surface ion trap can be mounted on top of the tungsten mounting structure for the lower Halbach array.  This component can then be placed in direct thermal contact with a cryogenic heat exchanger assembly if desired.
\end{subsection}

\begin{subsection}{Upper magnet array mount}
	The upper magnet array can be mounted in a titanium support block.  While titanium has a higher coefficient of thermal contraction than tungsten, it is far easier to machine, particularly to fine, sub-micron tolerances.  The use of titanium for this component is possible as the upper magnet block will not be in direct contact with the heat exchanger copper block, having several joints and a titanium to tungsten interface separating it from the cold-head of the heat exchanger.  The upper magnet array can thus be set in the correct configuration and bonded with either of the aforementioned methods, separated by titanium spacers machined to within $1$ $\mu$m of the desired magnet placement.  The upper array can be fixed in the correct position and mounted to the bracket support, discussed below, using a pair of threaded bolt-holes.
\end{subsection}	

\begin{subsection}{Bracket support}
	The two magnet arrays can be joined together, and the relative positions fixed by a titanium bracket support.  Once again, titanium is suggested as the part could be easily machined to very fine tolerance, in order to ensure that the relative positions of the magnet arrays can be set accurately and precisely, in accordance with the desirable configuration determined from simulations.  The titanium parts can be assembled with bolts, accepted by the threaded-holes, drilled in the correct mounting position during manufacturing.  It would also be possible to design this part with cut-outs for the bolts to allow the structure to be adjusted.  This may be desirable for fine-tuning the setup or changing the position based on further simulations for changes or improvements in the setup, but also introduces potential mechanical weaknesses which may fail under heating or cooling.  The bracket support could be mounted to the lower Halbach array, tungsten mounting block using cut bolt-holes or, if this is not possible due to machining issues with the tungsten, a titanium clamp or indium or epoxy bond is possible. 
\end{subsection}

\begin{section}{Modified Halbach array in a scalable architecture concept}
	\label{sec:MagnetScalability}
	The simulated results obtained for the Halbach magnet design suggest that a two-qubit gate scheme using long wavelength radiation may be effected with this system.  Following the design considerations of the previous, and as motivation for the proof-of-principle experiment, the implementation of such a permanent magnet design into a scalable, modular architecture is suggested, by modification of the particular QCCD concept presented by Lekitsch et al. \cite{Lekitsch2015}.
	
	To illustrate the integration of the Halbach array magnet design into a larger-scale device, a 2D array of nine x-junction ion traps is presented in figure \ref{fig:MagnetScalableConcept}.  The magnet arrays can be placed at the extremities of the device, with the upper-most and lower-most arms of the top and bottom rows of x-junctions respectively, used as quantum gate regions.  Ions can then be shuttled around the architecture arbitrarily by control of voltages to the DC electrodes fabricated on each of the surface ion traps in the array.  Quantum control of such single and two-qubit entangling gates can be applied in a designated gate zone, the region in which the strong gradient and field nil is present as a result of the modified Halbach arrays.  The magnet arrays, presented in \ref{fig:MagnetScalableConcept}, would be composed of the dual layer magnet array, presented in sections \ref{sec:RhombicCentreGeometry}, and could be mounted in a support structure as suggested above.  Ions could then be shuttled away from the gate regions, and find a path around the gate zones for continued propagation of computation for larger-scale operation.  If desired, the outer most central x-junction arms can also feature a magnet array, whereby the ions can be shuttled out of the featured nine junction module presented, using the four remaining unobstructed, external arms.  Subsequently connected nine junction arrays could then be tessellated such that the central magnet arrays are placed on alternating arms.  Redundant x-junctions, which do not feature gates regions, could also be placed between modules in a larger tessellation of modules, to facilitate shuttling connectivity between modules.  In a scalable system, small compensation of the magnetic field offsets could be effected using embedded solenoids or current-carrying wires, placed alongside the magnet arrays or micro-fabricated into the ion trap structure \cite{Bautista2015,Welzel2011, Kunert2014}.  The advantage of a combined approach featuring permanent magnets and solenoids is the minimization of heating and heat management associated with the large electrical currents required to generate a strong gradient using only an embedded wire.  The relatively modest current requirements for the purpose of compensating small offsets in the magnetic field, when the wires are placed close to the ion, should require very limited thermal management.  Another advantage is the simpler and more feasible engineering of ultra-stable current sources for such a task when the total power requirements are small, particularly in the case of increasing system scale.  A complete discussion on current carrying wires including analysis of current stability and heating for a given magnetic field are provided in the thesis of Dr F. R. Lebrun-Gallagher.
	\vskip4mm
	
	\begin{figure}[H]
		\centering
		\includegraphics[width=\linewidth]{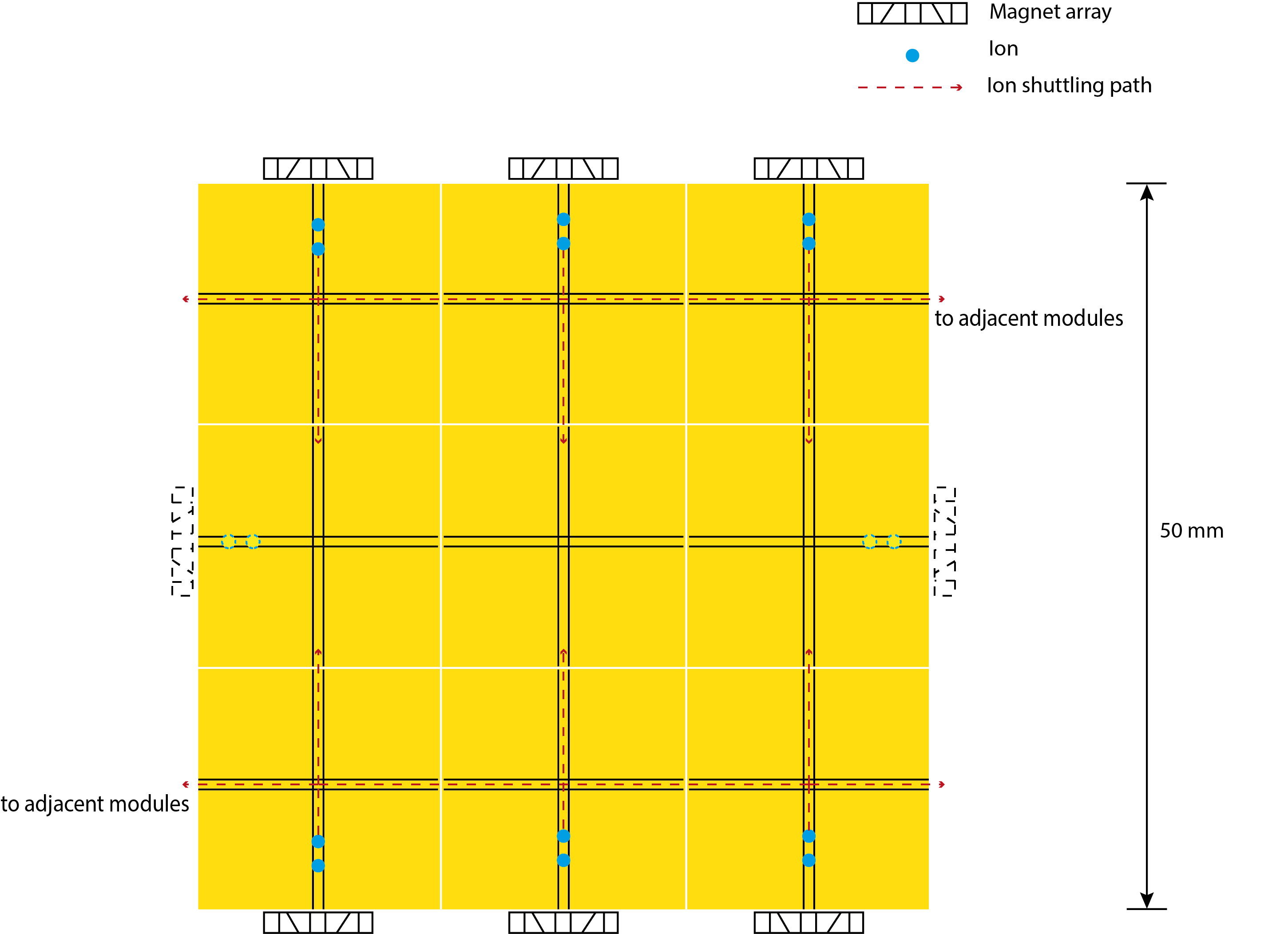}
		\caption{Concept diagram of an example repeating unit for implementing the magnet array design into a modular, scalable concept device similarly to the proposal by Lekitch et al. \cite{Lekitsch2015}.  The possible locations of magnet arrays, ions and shuttling are indicated.  The broken line magnet array locations represent optional placement of arrays, whereby access can be left open for ion transport to adjacent modules.}
		\label{fig:MagnetScalableConcept}
	\end{figure}
\end{section}

\section{Conclusion}
From the initial results, the achievable gradient at a minimum magnetic field offset in the axial direction is $99$ Tm$^{-1}$.  A small magnetic field contribution is apparent in the axial direction before dropping to the nil position.  This offset is still relatively small, however, and due to the fixed nature of the design, will induce a deterministic phase rotation on an ion qubit which is shuttled through the region at a fixed rate.  Furthermore, shuttling operations into and out of the region could be matched such that induced phase rotations into and out of the region are self-canceling.  It has been shown that other geometries are available with variations of the same format, which yield a smaller magnetic field offset in the region approaching the nil if required, at the cost of a reduced gradient.  Optimization of the desirable properties was attempted by exploring the parameter space of the design.  This process was enacted while accounting for the feasibility of building a scalable quantum computing architecture using the modified Halbach array paradigm.  The final, rhombic prism design was shown to offer a reduced, however still significant, magnetic field gradient, $\approx 51$ Tm$^{-1}$, while minimizing the absolute magnetic field in the ion transport region.  The asymmetry of the array geometry with respect to the $x,z$-plane means that even if magnets required for compensation are placed above the ion trap surface, they should not provide an obstruction for laser access, and allow the ions to be shuttled in and out of the gate region via the same path.  The negligible magnetic field at an extent of $\gtrsim 7$ mm from the magnets should allow ions to be shuttled around gate regions in a larger-scale device featuring multiple ion traps, such as that suggested by Lekitsch et al. \cite{Lekitsch2015} without significant perturbation.  Parameter optimizations choices were presented whereby constraints on the space of possible configurations were placed to serve a demonstrative proof-of-concept experiment.  It should be emphasized however, that the designs and simulations herein cannot be completely exhaustive, whereby a parameter space yet exists in which more optimal configurations and favorable properties may be exhibited dependent upon experiment design and engineering trade-offs when implementing experimentally or scaling the design to larger architectures.  This is particularly relevant for taking into account the complexity of a larger-scale system with multiple sets of magnet arrays in an inhomogeneous magnetic environment.  The limited resources for obtaining and/or cutting the magnets were taken into account here and used as a constraint to bound the parameter space, however, larger-scale production and manufacturing may yield design options with more favorable characteristics.  The particular design characteristics may also vary depending upon the peculiarities of a scalable system design, quantum gate implementation and system engineering features which provide better mitigation for magnetic field offset or phase accumulation due to the inhomogeneity in the field across the device.

\section*{Acknowledgments}

This work was supported by the U.K. Engineering and Physical Sciences Research Council via the EPSRC Hub in Quantum Computing and Simulation (EP/T001062/1), the European Commissions Horizon-2020 Flagship on Quantum Tehcnologies Project No. 820314 (MicroQC), the U.S. Army Research Office under Contract No. W911NF-14-2-0106 and Contract No. W911NF-21-1-0240, the Office of Naval Research under Agreement No. N62909-19-1-2116 and the University of Sussex.  The author would like to thank Prof. Winfried Hensinger for his guidance, support and discussions regarding this work and its potential application, and for feedback during the editing process for the manuscript.

\section*{Conflicts of Interest Statement}
The author has no conflicts of interest, financial or commercial.

\newpage

\bibliography{magnet_paper}

@article{Lekitsch2015,
author = {Bjoern Lekitsch  and Sebastian Weidt  and Austin G. Fowler  and Klaus Mølmer  and Simon J. Devitt  and Christof Wunderlich  and Winfried K. Hensinger },
title = {Blueprint for a microwave trapped ion quantum computer},
journal = {Science Advances},
volume = {3},
number = {2},
pages = {e1601540},
year = {2017},
doi = {10.1126/sciadv.1601540},
URL = {https://www.science.org/doi/abs/10.1126/sciadv.1601540},
eprint = {https://www.science.org/doi/pdf/10.1126/sciadv.1601540}}

@phdthesis{Murgia2017,
	author = {Murgia, D. F.},
	school = {Imperial College London},
	title = {{Microchip ion traps with high magnetic field gradients for microwave quantum logic}},
	year = {2017}
}

@article{Johanning2009,
	archivePrefix = {},
	arxivId = {},
	author = {Johanning, M. and Braun, A. and Timoney, N. and Elman, V. and Neuhauser, W. and Wunderlich, Chr},
	doi = {10.1103/PhysRevLett.102.073004},
	eprint = {0801.0078},
	isbn = {0031-9007 (Print)$\backslash$r0031-9007 (Linking)},
	issn = {10797114},
	journal = {Physical Review Letters},
	number = {7},
	pages = {1--4},
	pmid = {19257664},
	title = {{Individual addressing of trapped ions and coupling of motional and spin states using rf radiation}},
	volume = {102},
	year = {2009}
}

@article{Mintert2001,
	archivePrefix = {arXiv},
	arxivId = {quant-ph/0104041},
	author = {Mintert, F. and Wunderlich, C.},
	doi = {10.1103/PhysRevLett.87.257904},
	eprint = {0104041},
	isbn = {0031-9007},
	issn = {10797114},
	journal = {Physical Review Letters},
	number = {25},
	pages = {257904--1--257904--4},
	pmid = {11736608},
	primaryClass = {quant-ph},
	title = {{Ion-trap quantum logic using long-wavelength radiation}},
	volume = {87},
	year = {2001}
}

@article{Lake2015,
	archivePrefix = {arXiv},
	arxivId = {1409.1862},
	author = {Lake, K. and Weidt, S. and Randall, J. and Standing, E. D. and Webster, S. C. and Hensinger, W. K.},
	doi = {10.1103/PhysRevA.91.012319},
	eprint = {1409.1862},
	issn = {10941622},
	journal = {Physical Review A - Atomic, Molecular, and Optical Physics},
	number = {1},
	pages = {1--5},
	title = {{Generation of spin-motion entanglement in a trapped ion using long-wavelength radiation}},
	volume = {91},
	year = {2015}
}

@article{Weidt2016,
  title = {Trapped-Ion Quantum Logic with Global Radiation Fields},
  author = {Weidt, S. and Randall, J. and Webster, S. C. and Lake, K. and Webb, A. E. and Cohen, I. and Navickas, T. and Lekitsch, B. and Retzker, A. and Hensinger, W. K.},
  journal = {Phys. Rev. Lett.},
  volume = {117},
  issue = {22},
  pages = {220501},
  numpages = {6},
  year = {2016},
  month = {Nov},
  publisher = {American Physical Society},
  doi = {10.1103/PhysRevLett.117.220501},
  url = {https://link.aps.org/doi/10.1103/PhysRevLett.117.220501}
}

@article{Harty2014,
	author = {Harty, T. P. and Allcock, D. T.C. and Ballance, C. J. and Guidoni, L. and Janacek, H. A. and Linke, N. M. and Stacey, D. N. and Lucas, D. M.},
	title = {High-fidelity preparation, gates, memory, and readout of a trapped-ion quantum bit},
	arxivId = {1403.1524},
	doi = {10.1103/PhysRevLett.113.220501},
	eprint = {1403.1524},
	issn = {10797114},
	journal = {Physical Review Letters},
	number = {22},
	pages = {2--6},
	volume = {113},
	year = {2014}
}

@article{Ballance2016,
	title = {High-Fidelity Quantum Logic Gates Using Trapped-Ion Hyperfine Qubits},
	author = {Ballance, C. J. and Harty, T. P. and Linke, N. M. and Sepiol, M. A. and Lucas, D. M.},
	journal = {Phys. Rev. Lett.},
	volume = {117},
	issue = {6},
	pages = {060504},
	numpages = {6},
	year = {2016},
	month = {Aug},
	publisher = {American Physical Society},
	doi = {10.1103/PhysRevLett.117.060504},
	url = {https://link.aps.org/doi/10.1103/PhysRevLett.117.060504}
}

@article{Kjaergaard2020,
	author = {Kjaergaard, M. and Schwartz, M. E. and Braumüller, J. and Krantz, P. and Wang, J. I.-J. and Gustavsson, S. and Oliver, W. D.},
	title = {Superconducting Qubits: Current State of Play},
	journal = {Annual Review of Condensed Matter Physics},
	volume = {11},
	number = {1},
	pages = {369-395},
	year = {2020},
	doi = {10.1146/annurev-conmatphys-031119-050605},
	
	URL = { 
	https://doi.org/10.1146/annurev-conmatphys-031119-050605
	
	},
	eprint = { 
	https://doi.org/10.1146/annurev-conmatphys-031119-050605
	
	}
}

@article{Stephenson2020,
	title = {High-Rate, High-Fidelity Entanglement of Qubits Across an Elementary Quantum Network},
	author = {Stephenson, L. J. and Nadlinger, D. P. and Nichol, B. C. and An, S. and Drmota, P. and Ballance, T. G. and Thirumalai, K. and Goodwin, J. F. and Lucas, D. M. and Ballance, C. J.},
	journal = {Phys. Rev. Lett.},
	volume = {124},
	issue = {11},
	pages = {110501},
	numpages = {6},
	year = {2020},
	month = {Mar},
	publisher = {American Physical Society},
	doi = {10.1103/PhysRevLett.124.110501},
	url = {https://link.aps.org/doi/10.1103/PhysRevLett.124.110501}
}

@article{Kawai2016,
	doi = {10.1088/1361-6455/50/2/025501},
	url = {https://doi.org/10.1088/1361-6455/50/2/025501},
	year = 2016,
	month = {dec},
	publisher = {{IOP} Publishing},
	volume = {50},
	number = {2},
	pages = {025501},
	author = {Yuji, K. and Kenji, S. and Atsushi, N. and Shinji, U. and Utako, T.},
	title = {Surface-electrode trap with an integrated permanent magnet for generating a magnetic-field gradient at trapped ions},
	journal = {Journal of Physics B: Atomic, Molecular and Optical Physics}
}

@article{Hucul2008,
author = {Hucul, D. and Yeo, M. and Olmschenk, S. and Monroe, C. and Hensinger, W. K. and Rabchuk, J.},
title = {On the transport of atomic ions in linear and multidimensional ion trap arrays},
year = {2008},
issue_date = {July 2008},
publisher = {Rinton Press, Incorporated},
address = {Paramus, NJ},
volume = {8},
number = {6},
issn = {1533-7146},
journal = {Quantum Info. Comput.},
month = jul,
pages = {501–578},
numpages = {78}
}

@article{Monroe2014,
	title = {Large-scale modular quantum-computer architecture with atomic memory and photonic interconnects},
	author = {Monroe, C. and Raussendorf, R. and Ruthven, A. and Brown, K. R. and Maunz, P. and Duan, L.-M. and Kim, J.},
	journal = {Phys. Rev. A},
	volume = {89},
	issue = {2},
	pages = {022317},
	numpages = {16},
	year = {2014},
	month = {Feb},
	publisher = {American Physical Society},
	doi = {10.1103/PhysRevA.89.022317},
	url = {https://link.aps.org/doi/10.1103/PhysRevA.89.022317}
}

@article{Kaushal2020,
	author = {Kaushal, V.  and Lekitsch, B.  and Stahl, A.  and Hilder, J.  and Pijn, D.  and Schmiegelow, C.  and Bermudez, A.  and Müller, M.  and Schmidt-Kaler, F.  and Poschinger, U. },
	title = {Shuttling-based trapped-ion quantum information processing},
	journal = {AVS Quantum Science},
	volume = {2},
	number = {1},
	pages = {014101},
	year = {2020},
	doi = {10.1116/1.5126186},
	
	URL = { 
	https://doi.org/10.1116/1.5126186
	
	},
	eprint = { 
	https://doi.org/10.1116/1.5126186
	
	}
}

@article{Kunert2014,
	author = {Kunert, P. and Georgen, D. and Bogunia, L. and Baig, Muhammad T. and Baggash, M. and Johanning, M. and Wunderlich, C.},
	year = {2014},
	month = {01},
	pages = {27-36},
	title = {A planar ion trap chip with integrated structures for an adjustable magnetic field gradient},
	volume = {114},
	journal = {Applied Physics B},
	doi = {10.1007/s00340-013-5722-9}
}

@Article{Welzel2011,
author={Welzel, J.
and Bautista-Salvador, A.
and Abarbanel, C.
and Wineman-Fisher, V.
and Wunderlich, C.
and Folman, R.
and Schmidt-Kaler, F.},
title={Designing spin-spin interactions with one and two dimensional ion crystals in planar micro traps},
journal={The European Physical Journal D},
year={2011},
month={Nov},
day={01},
volume={65},
number={1},
pages={285-297},
issn={1434-6079},
doi={10.1140/epjd/e2011-20098-y},
url={https://doi.org/10.1140/epjd/e2011-20098-y}
}

@phdthesis{Bautista2015,
	author = {Bautista-Salvador, A.},
	title = {{Integrated Electromagnets and Radiofrequency Spectroscopy in a Planar Paul Trap}},
	year = {2015},
    school = {Universität Ulm
Institut für Quanteninformationsverarbeitung, Johannes Gutenberg-Universität Mainz
Institut für Physik}
}

@Article{Akhtar2022,
author={Akhtar, M.
and Bonus, F.
and Lebrun-Gallagher, F. R.
and Johnson, N. I.
and Siegele-Brown, M.
and Hong, S.
and Hile, S. J.
and Kulmiya, S. A.
and Weidt, S.
and Hensinger, W. K.},
title={A high-fidelity quantum matter-link between ion-trap microchip modules},
journal={Nature Communications},
year={2023},
month={Feb},
day={08},
volume={14},
number={1},
pages={531},
issn={2041-1723},
doi={10.1038/s41467-022-35285-3},
url={https://doi.org/10.1038/s41467-022-35285-3}
}

@article{Siegele-Brown_2022,
	doi = {10.1088/2058-9565/ac66fc},
	url = {https://dx.doi.org/10.1088/2058-9565/ac66fc},
	year = {2022},
	month = {May},
	publisher = {IOP Publishing},
	volume = {7},
	number = {3},
	pages = {034003},
	author = {M. Siegele-Brown and S. Hong and F. R. Lebrun-Gallagher and S. J. Hile and S. Weidt and W. K. Hensinger},
	title = {Fabrication of surface ion traps with integrated current carrying wires enabling high magnetic field gradients},
	journal = {Quantum Science and Technology}
}

@phdthesis{Peaks2023,
    author = {Peaks, M. G.},
    title = {Strong magnetic field gradients for scalable trapped ion quantum logic},
    school = {University of Sussex},
    year = {2023},
    month = {9},
    url = {https://hdl.handle.net/10779/uos.24057888.v1}
}

@article{Bruzewicz2019Review,
	author = {Bruzewicz, C. D. and Chiaverini, J. and McConnell, R. and Sage, J. M.},
	title = {Trapped-ion quantum computing: Progress and challenges},
	journal = {Applied Physics Reviews},
	volume = {6},
	number = {2},
	pages = {021314},
	year = {2019},
	month = {05},
	issn = {1931-9401},
	doi = {10.1063/1.5088164},
	url = {https://doi.org/10.1063/1.5088164},
	eprint = {https://pubs.aip.org/aip/apr/article-pdf/doi/10.1063/1.5088164/14577412/021314\_1\_online.pdf}
}

@article{Henriet2020quantumcomputing,
	doi = {10.22331/q-2020-09-21-327},
	url = {https://doi.org/10.22331/q-2020-09-21-327},
	title = {Quantum computing with neutral atoms},
	author = {Henriet, L. and Beguin, L. and Signoles, A. and Lahaye, T. and Browaeys, A. and Reymond, G. and Jurczak, C.},
	journal = {{Quantum}},
	issn = {2521-327X},
	publisher = {{Verein zur F{\"{o}}rderung des Open Access Publizierens in den Quantenwissenschaften}},
	volume = {4},
	pages = {327},
	month = {Sep},
	year = {2020}
}

@article{Monroe2013,
	author = {C. Monroe  and J. Kim},
	title = {Scaling the Ion Trap Quantum Processor},
	journal = {Science},
	volume = {339},
	number = {6124},
	pages = {1164-1169},
	year = {2013},
	doi = {10.1126/science.1231298},
	URL = {https://www.science.org/doi/abs/10.1126/science.1231298},
	eprint = {https://www.science.org/doi/pdf/10.1126/science.1231298}}

@article{Wineland1998,
	author = {D. J. Wineland and C. Monroe},
	title = {Experimental Issues in Coherent Quantum-State Manipulation of Trapped Atomic Ions},
	journal = {Journal of research of the National Institute of Standards and Technology},
	volume = {103},
	number = {3},
	pages = {259-328},
	year = {1998},
	doi = {10.6028/jres.103.019},
	URL = {https://pubmed.ncbi.nlm.nih.gov/28009379/}
}

@Article{Wang2021,
author={Wang, P. 
and Luan, C.
and Qiao, M.
and Um, M.
and Zhang, J.
and Wang, Y.
and Yuan, X.
and Gu, M.
and Zhang, J.
and Kim, K.},
title={Single ion qubit with estimated coherence time exceeding one hour},
journal={Nature Communications},
year={2021},
month={Jan},
day={11},
volume={12},
number={1},
pages={233},
issn={2041-1723},
doi={10.1038/s41467-020-20330-w},
url={https://doi.org/10.1038/s41467-020-20330-w}
}

@article{Heinzen1990,
	title = {Quantum-limited cooling and detection of radio-frequency oscillations by laser-cooled ions},
	author = {Heinzen, D. J. and Wineland, D. J.},
	journal = {Phys. Rev. A},
	volume = {42},
	issue = {5},
	pages = {2977--2994},
	numpages = {0},
	year = {1990},
	month = {Sep},
	publisher = {American Physical Society},
	doi = {10.1103/PhysRevA.42.2977},
	url = {https://link.aps.org/doi/10.1103/PhysRevA.42.2977}
	}

@article{Srinivas2021,
author={Srinivas, R.
and Burd, S. C.
and Knaack, H. M.
and Sutherland, R. T.
and Kwiatkowski, A.
and Glancy, S.
and Knill, E.
and Wineland, D. J.
and Leibfried, D.
and Wilson, A. C.
and Allcock, D. T. C.
and Slichter, D. H.},
title={High-fidelity laser-free universal control of trapped ion qubits},
journal={Nature},
year={2021},
month={Sep},
day={01},
volume={597},
number={7875},
pages={209-213},
issn={1476-4687},
doi={10.1038/s41586-021-03809-4},
url={https://doi.org/10.1038/s41586-021-03809-4}
}

@article{Ospelkaus2008,
	author = {C. Ospelkaus and C. E. Langer and J. M. Amini and K. R. Brown and D. Leibfried and D. J. Wineland},
	doi = {10.1103/PhysRevLett.101.090502},
	issn = {00319007},
	issue = {9},
	journal = {Physical Review Letters},
	month = {8},
	title = {Trapped-ion quantum logic gates based on oscillating magnetic fields},
	volume = {101},
	year = {2008},
}

@article{Ozeri2007,
	title = {Errors in trapped-ion quantum gates due to spontaneous photon scattering},
	author = {Ozeri, R. and Itano, W. M. and Blakestad, R. B. and Britton, J. and Chiaverini, J. and Jost, J. D. and Langer, C. and Leibfried, D. and Reichle, R. and Seidelin, S. and Wesenberg, J. H. and Wineland, D. J.},
	journal = {Phys. Rev. A},
	volume = {75},
	issue = {4},
	pages = {042329},
	numpages = {14},
	year = {2007},
	month = {Apr},
	publisher = {American Physical Society},
	doi = {10.1103/PhysRevA.75.042329},
	url = {https://link.aps.org/doi/10.1103/PhysRevA.75.042329}
}

@article{Gidney2021,
	title={How to factor 2048 bit RSA integers in 8 hours using 20 million noisy qubits},
	author={Gidney, C. and Eker{\aa}, M.},
	journal={Quantum},
	volume={5},
	pages={433},
	year={2021},
	publisher={Verein zur F{\"o}rderung des Open Access Publizierens in den Quantenwissenschaften}
}

@article{Kivlichan2020,
	title={Improved fault-tolerant quantum simulation of condensed-phase correlated electrons via trotterization},
	author={Kivlichan, I. D and Gidney, C. and Berry, D. W and Wiebe, N. and McClean, J. and Sun, W. and Jiang, Z. and Rubin, N. and Fowler, A. and Aspuru-Guzik, A. and others},
	journal={Quantum},
	volume={4},
	pages={296},
	year={2020},
	publisher={Verein zur F{\"o}rderung des Open Access Publizierens in den Quantenwissenschaften}
}

@article{Reiher2017,
	author = {M. Reiher  and N. Wiebe  and K. M. Svore  and D. Wecker  and M. Troyer },
	title = {Elucidating reaction mechanisms on quantum computers},
	journal = {Proceedings of the National Academy of Sciences},
	volume = {114},
	number = {29},
	pages = {7555-7560},
	year = {2017},
	doi = {10.1073/pnas.1619152114},
	URL = {https://www.pnas.org/doi/abs/10.1073/pnas.1619152114},
	eprint = {https://www.pnas.org/doi/pdf/10.1073/pnas.1619152114}
}

@article{Egan2021,
	title={Fault-tolerant control of an error-corrected qubit},
	author={Egan, L. and Debroy, D. M and Noel, C. and Risinger, A. and Zhu, D. and Biswas, D. and Newman, M. and Li, M. and Brown, K. R. and Cetina, M. and others},
	journal={Nature},
	volume={598},
	number={7880},
	pages={281--286},
	year={2021},
	publisher={Nature Publishing Group UK London}
}

@article{Ryan2021,
	title={Realization of real-time fault-tolerant quantum error correction},
	author={Ryan-Anderson, C. and Bohnet, J. G. and Lee, K. and Gresh, D. and Hankin, A. and Gaebler, J. P. and Francois, D. and Chernoguzov, A. and Lucchetti, D. and Brown, N. C. and others},
	journal={Physical Review X},
	volume={11},
	number={4},
	pages={041058},
	year={2021},
	publisher={APS}
}

@article{Krinner2022,
	title={Realizing repeated quantum error correction in a distance-three surface code},
	author={Krinner, S. and Lacroix, N. and Remm, A. and Di Paolo, A. and Genois, E. and Leroux, C. and Hellings, C. and Lazar, S. and Swiadek, F. and Herrmann, J. and others},
	journal={Nature},
	volume={605},
	number={7911},
	pages={669--674},
	year={2022},
	publisher={Nature Publishing Group UK London}
}

@article{Zhao2022,
	title={Realization of an error-correcting surface code with superconducting qubits},
	author={Zhao, Y. and Ye, Y. and Huang, H. and Zhang, Y. and Wu, D. and Guan, H. and Zhu, Q. and Wei, Z. and He, T. and Cao, S. and others},
	journal={Physical Review Letters},
	volume={129},
	number={3},
	pages={030501},
	year={2022},
	publisher={APS}
}

@article{Wineland1997,
author = {Wineland, D. and Monroe, C. and Itano, Wayne and Leibfried, D. and King, B. and Meekhof, D.},
year = {1997},
month = {11},
pages = {},
title = {Experimental Issues in Coherent Quantum-State Manipulation of Trapped Atomic Ions},
volume = {103},
journal = {Journal of Research of the National Institute of Standards and Technology},
doi = {10.6028/jres.103.019}
}

@article{Monroe_Raman1995,
	title = {Resolved-Sideband Raman Cooling of a Bound Atom to the 3D Zero-Point Energy},
	author = {Monroe, C. and Meekhof, D. M. and King, B. E. and Jefferts, S. R. and Itano, W. M. and Wineland, D. J. and Gould, P.},
	journal = {Phys. Rev. Lett.},
	volume = {75},
	issue = {22},
	pages = {4011--4014},
	numpages = {0},
	year = {1995},
	month = {Nov},
	publisher = {American Physical Society},
	doi = {10.1103/PhysRevLett.75.4011},
	url = {https://link.aps.org/doi/10.1103/PhysRevLett.75.4011}
}

@Article{Debnath2016,
	author={Debnath, S.
	and Linke, N. M.
	and Figgatt, C.
	and Landsman, K. A.
	and Wright, K.
	and Monroe, C.},
	title={Demonstration of a small programmable quantum computer with atomic qubits},
	journal={Nature},
	year={2016},
	month={Aug},
	day={01},
	volume={536},
	number={7614},
	pages={63-66},
	issn={1476-4687},
	doi={10.1038/nature18648},
	url={https://doi.org/10.1038/nature18648}
}

\end{document}